\documentclass[12pt]{article}
\usepackage{latexsym}
\usepackage{epsfig,amssymb,euscript}
\usepackage{amsmath}
\usepackage{mathrsfs}
\def\be{\begin{equation}}
\def\ee{\end{equation}}
\def\bea{\begin{eqnarray}}
\def\eea{\end{eqnarray}}

\numberwithin{equation}{section}       

\newcommand{\nn}{\nonumber}

\newcommand\B{\mathbb{B}}
\newcommand\R{\mathbb{R}}
\newcommand\Z{\mathbb{Z}}

\newcommand\N{\mathbb{N}}

\newcommand{\de}{\partial}

\newcommand{\vol}{\mathrm{vol}}
\newcommand{\ai}{a_i}
\def\aone{a_1}
\def\atwo{a_2}
\def\athree{a_3}

\begin{document}
\begin{titlepage}
\begin{center}
\today
\vspace*{1.8cm}

{\Large \bf Fivebranes and resolved deformed $G_2$ manifolds}

\vskip 1.7cm
J\'er\^ome Gaillard$^{1}$ and  Dario Martelli$^{2}$
\vskip 1.5cm

{$^1$\em  Swansea University,\\
Singleton Park, Swansea SA2 8PP, United Kingdom}\\

\vskip 1.5cm

{$^2$ \em  Department of Mathematics, King's College, London\\
The Strand, London WC2R 2LS, United Kingdom}

\end{center}

\vskip 3.5 cm
\begin{abstract}
\noindent
We study supergravity  solutions  corresponding to fivebranes wrapped on a three-sphere inside
a $G_2$ holonomy manifold. By changing a parameter the solutions interpolate between a $G_2$ manifold 
$X_i\cong S^3\times \R^4$ with flux on a three-sphere and a distinct $G_2$ manifold $X_j\cong S^3\times \R^4$ 
with branes on another three-sphere. Therefore, these realise  a $G_2$  geometric transition purely in the
supergravity context. We can add D2 brane charge by applying a simple transformation to the initial solution
and we obtain one-parameter deformations of warped $G_2$ holonomy backgrounds. These solutions
suggest a connection between the ${\cal N}=1$ Chern-Simons theory on the fivebranes and the field theory dual
to D2 branes and fractional NS5 branes, transverse to the  $G_2$ manifold $S^3\times \R^4$.

\end{abstract}

\end{titlepage}
\pagestyle{plain}
\setcounter{page}{1}
\newcounter{bean}
\baselineskip18pt
\tableofcontents

\section{Introduction}

Calabi-Yau manifolds have played an important role in string theory for many years. 
For example,  string theory compactified on a
Calabi-Yau manifold gives rise to an effective four-dimensional theory with unbroken supersymmetry. 
In the context of the AdS/CFT correspondence 
\cite{Maldacena:1997re}, conical Calabi-Yau three-fold singularities equipped with a 
Ricci flat metric give rise to supersymmetric solutions of Type IIB string theory with an AdS$_5$ factor
\cite{Kehagias:1998gn,Klebanov:1998hh,Acharya:1998db,Morrison:1998cs}, 
thus providing precise gravity duals to a large class of supersymmetric field theories.

The simplest and most studied Calabi-Yau singularity in string theory is the conifold \cite{Candelas:1989js}. 
There are two distinct desingularisations of this, in which the singularity at the tip 
 is replaced by a three-sphere or a two-sphere.
These  are referred to as \emph{deformation} and \emph{resolution}, respectively. 
The relevance of the transition between the resolved and deformed conifold in string theory was 
emphasised in \cite{Strominger:1995cz}.  The deformed conifold geometry underlies the Klebanov-Strassler solution 
\cite{Klebanov:2000hb}, which is dual to a four-dimensional ${\cal N}=1$ field theory displaying confinement. 
A different supergravity solution dual to a closely related field theory  was discussed by Maldacena-Nu\~nez 
\cite{Maldacena:2000yy,Chamseddine:1997mc}. This arises as the decoupling limit of a configuration of fivebranes wrapped on the 
two-sphere of the resolved conifold.

A solution of Type IIB supergravity that contains as special cases the Klebanov-Strassler and the Maldacena-Nu\~nez solutions
was constructed in \cite{Butti:2004pk}. This was interpreted as the gravity dual to 
the baryonic branch of the Klebanov-Strassler theory, with a non-trivial parameter related to
the VEV of the baryonic operators \cite{Gubser:2004qj,Dymarsky:2005xt}. 
It was later pointed out in \cite{Maldacena:2009mw} 
that the solution of \cite{Butti:2004pk} is related to a simpler solution \cite{Casero:2006pt}  corresponding 
to fivebranes wrapped on the two-sphere of the resolved conifold, without taking any near-brane limit. 
In this context, the non-trivial parameter can roughly be viewed as the size of the two-sphere wrapped by the branes. When this is very
large,  the solution looks like the resolved conifold with branes on the two-sphere 
and when this is very small, it looks like the deformed conifold with flux 
on the three-sphere. Therefore, it displays an explicit realisation of the geometric transition described in \cite{Vafa:2000wi}.
The key fact that allows the connection of the resolved and
deformed conifold at the classical level is that the solution is an example of non-K\"ahler, or \emph{torsional}, geometry  \cite{Strominger:1986uh,Hull:1986kz}. For related work see \cite{Halmagyi:2010st}.

In this paper we will present a $G_2$ version of the picture advocated in \cite{Maldacena:2009mw}. 
In particular, we will discuss  supergravity solutions which correspond to $M$ 
fivebranes wrapped on the three-sphere inside the $G_2$ holonomy manifold
$S^3\times \R^4$ \cite{Bryand:1989mv,Gibbons:1989er}, without taking any near-brane limit. 
Solutions of this type were previously discussed in \cite{Canoura:2008at}.
If we take the near-brane limit, we find the solutions discussed 
by Maldacena-Nastase in \cite{Maldacena:2001pb}, based on \cite{Chamseddine:2001hk}. These were argued in 
\cite{Maldacena:2001pb} to be the gravity dual of ${\cal N}=1$ supersymmetric $U(M)$ 
Chern-Simons theories in three dimensions, 
with Chern-Simons level $|k|=M/2$.

The (classical) moduli space of asymptotically conical $G_2$ holonomy metrics on $S^3\times \R^4$ 
comprises \emph{three} branches, that we will denote $X_i$ 
\cite{Atiyah:2001qf},  intersecting on the singular cone.
These three branches are  related to the 
three branches of the conifold moduli space, namely the deformation, 
the resolution and the flopped resolution \cite{Atiyah:2000zz,Atiyah:2001qf}.
One therefore expects close analogies with the discussion in \cite{Maldacena:2009mw}. 
Indeed, by working in the context of 
\emph{torsional $G_2$ manifolds} \cite{Gauntlett:2002sc}, 
we will find a set of one-parameter families of solutions that pairwise 
interpolate between the three classical branches of $G_2$ holonomy.  
The  non-trivial parameter in these  solutions 
can roughly be viewed as the size of the three-sphere wrapped by the fivebranes. 
When this is very large, the solution looks like a $G_2$ manifold  $X_i$
with branes on a three-sphere 
and when this is very small, it looks like a distinct $G_2$ manifold  $X_j$ ($i\neq j$)
with flux on a different three-sphere. This then  realises  a \emph{$G_2$ geometric transition}.
However, we will not attempt to relate this to a ``large $N$ duality'' as in the original 
discussion in \cite{Vafa:2000wi}. More concretely, we find \emph{six} distinct solutions
connecting the three classical branches $X_i$, that we will denote $X_{ij}$. 
In contrast to the conifold case, we can go from 
$X_i$ with branes to $X_j$ with flux or, conversely, 
from $X_j$ with branes to $X_i$ with flux, hence $X_{ij}\neq X_{ji}$. 
A rather different connection of the three classical  branches was discussed in \cite{Atiyah:2001qf}, 
where it was related to quantum effects in M-theory. 
Our discussion, on the other hand, is purely classical and ten-dimensional.

Starting from these relatively simple solutions, we will  
construct Type IIA  solutions with D2 brane charge and RR $C_3$ field. 
This can be done by applying a simple transformation analogous to the one discussed in 
\cite{Maldacena:2009mw} and further studied  in \cite{Minasian:2009rn,Gaillard:2010qg}. 
The solutions that we obtain in this way have 
 a warp factor (in the string frame metric) that becomes constant at infinity, 
 thus the geometry merges into an ordinary 
$G_2$ holonomy manifold. By taking a scaling limit we obtain solutions that become asymptotically
AdS$_4\times S^3\times S^3$, albeit only in the string frame. If we further tune the non-trivial parameter, 
we recover the backgrounds of \cite{Cvetic:2000mh}, corresponding 
to D2 branes and fractional NS5 branes transverse to the $G_2$ manifold $S^3\times \R^4$. 
Thus, our solutions may be thought of as  one-parameter deformations of the latter
and are analogous to the baryonic branch deformation 
\cite{Butti:2004pk}  of the Klebanov-Strassler geometry \cite{Klebanov:2000hb}. 
It is therefore very tempting to think  that there should be 
a close relation between the supersymmetric Chern-Simons 
theory discussed in \cite{Maldacena:2001pb} and the three-dimensional field theory on the D2 branes. 
We will  make some speculative considerations in the final section, but we will leave 
the field theory dual interpretation of our solutions for future work.

\section{Review of the $G_2$ holonomy manifold $S^3\times \R^4$}
\label{g2review}

In this section we review some aspects of the $G_2$ holonomy manifold $S^3\times \R^4$
that will be relevant for our discussion. We follow closely the presentation in \cite{Atiyah:2001qf}.

\subsection{The manifold and its topology}

The non-compact  seven-dimensional manifold defined by 
\bea
X =  \{ x_1^2 +x_2^2 +x_3^2 +x_4^2 - y_1^2 - y_2^2 -y_3^2 -y_4^2 =\epsilon~, ~ \quad \, x_i, y_i, \epsilon  \in \R \}
\eea
is the spin bundle over $S^3$ and is topologically equivalent to the manifold $S^3\times \R^4$.
 For $\epsilon>0$ the $S^3$ corresponds to the locus $y_i=0$ and the coordinates $y_i$
 parameterise the normal $\R^4$ directions.
This manifold admits  a Ricci-flat metric with $G_2$ holonomy, that at infinity approaches the cone metric
 \bea
 ds^2_{\mathrm{cone}} = dt^2 + t^2 ds^2 (Y)
 \eea
 where $Y\cong S^3 \times S^3$. The Einstein metric $ds^2(Y)$ is not the product of round metrics on the two three-spheres. 
It is in fact a nearly K\"ahler metric, which may be described in terms 
 of $SU(2)$ group elements as follows \cite{Atiyah:2001qf}. Consider three elements $\ai \in SU(2)$
 obeying the constraint
 \bea
 \aone \atwo \athree  = 1 ~.
 \label{constr}
 \eea 
There is an $SU(2)^3$ action preserving this relation given by $\ai \to  u_{i+1} \ai u_{i-1}^{-1}$, with $u_i\in SU(2)$, 
where the index $i$ is defined mod $3$. There is also an action by a ``triality''
group $\Sigma_3$, which is isomorphic to the group of permutations of three elements. 
This is an outer automorphism of the group $SU(2)^3$ and may be  
generated by\footnote{In the notation  of \cite{Atiyah:2001qf}  the generators of the group  $\Sigma_3$
 are denoted as $\sigma_{31} =\alpha$ and $\sigma_{231}=\beta$.}
\be
\begin{aligned}
\sigma_{31} &:\,   (\aone,\atwo,\athree) \, \rightarrow \, (\athree^{-1},\atwo^{-1},\aone^{-1}) ~,\\
\sigma_{231}  & : \, (\aone,\atwo,\athree) \, \rightarrow \, (\atwo,\athree,\aone)~.
\label{defs3}
\end{aligned}
\ee
The full list of group elements is given by $\Sigma_3 = \{e=\sigma_{123}, \sigma_{231}, \sigma_{312},\sigma_{12}, \sigma_{31}, \sigma_{23}  \}$,  
 with actions on $a_i$ following from \eqref{defs3}.

There are three different  seven-manifolds $X_1$, $X_2$, $X_3$, all homeomorphic to $S^3\times \R^4$, 
which can be obtained smoothing out the cone singularity by  blowing up three different three-spheres inside $Y$. These are permuted among each other by the action of $\Sigma_3$.  
This can be seen from the description of the base $Y\cong S^3\times S^3$ in terms of triples of group elements $(g_1, g_2, g_3) \in SU(2)^3$ subject to an equivalence 
relation $g_i \cong g_i h$ with $h\in SU(2)$, and is related to the previous description by setting $\ai = g_{i+1}g^{-1}_{i-1}$. 
We can  consider three different compact seven-manifolds $X_i'$, bounded by $Y$, obtained in each case by allowing  $g_i$ to take values in the four-ball $\B^4$. The non-compact seven-manifolds obtained after 
omitting the boundary are precisely the $X_i$. By setting $h=g_{i-1}^{-1}$, we see that each 
$X_i$ has   topology $S^3\times \R^4$, where 
$S^3$ and $\R^4$ are parameterised by  $g_{i+1}$ and $g_i$, respectively. 
We will review explicit metrics with $G_2$ holonomy in the three different cases in the next subsection. 

We can define three \emph{sub-manifolds} of $Y$ as
\bea
C_i = \{ a_i=1 \} \cong S^3
\label{subma}
\eea
which also extend to sub-manifolds in $X_i$, defined at some constant $t$.  
 These are topologically three-spheres, but
as cycles in $Y$ and $X_i$ they are not independent since the third Betti numbers of these manifolds
are $b_3(Y)=2$ and $b_3(X_i)=1$, respectively.  In fact, we have the following homology relation 
\bea
[C_1] + [C_2] + [C_3] = 0 \quad \mathrm{in}~~Y.
\eea
As cycles in $X_i$, the $[C_i]$ must obey an additional relation, which in view of their construction above is simply given by  $[C_i]=0$ in $X_i$. Therefore the third homology group  $H_3(X_i;\Z)$ is generated
by $C_{i-1}$ or $C_{i+1}$, with $[C_{i-1}]=-[C_{i+1}]$.

\subsection{A triality of $G_2$ holonomy metrics}
\label{g2metrics:section}

A seven-dimensional manifold is said to be a $G_2$ holonomy manifold if the holonomy 
group of the Levi-Civita connection $\nabla$
is contained in  $G_2\subset SO(7)$. It is well-known that these are characterised
by the existence of  a $G_2$ invariant three-form $\phi$ (associative three-form), together with its Hodge dual $*\phi$, which are both closed:
\bea
\mathrm{hol}(\nabla) \subseteq G_2 ~~~~~ \mathrm{iff} ~~~~~ d\phi = d* \phi = 0 ~.
\eea 
The metric compatible with these is Ricci-flat and  there exists a covariantly constant spinor, 
$\nabla \eta =0$. The $G_2$ invariant forms can be constructed from 
the constant spinor as bi-linears $\phi_{abc}= \eta^T \gamma_{abc}\eta$, $*\phi_{abcd}= \eta^T \gamma_{abcd}\eta$,  
and the metric is then uniquely determined by these. 
More generally, the two invariant forms define a $G_2$ \emph{structure} 
on the seven-dimensional manifold.  See for example  \cite{Gauntlett:2002sc}.

\begin{figure}[ht!]
\begin{center}
	\includegraphics[width=0.3\textwidth]{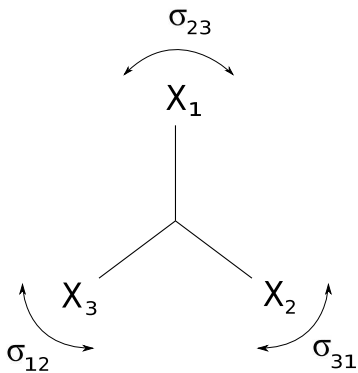} 
		\caption{The moduli space of asymptotically conical $G_2$ holonomy metrics on $S^3\times \R^4$ has three branches permuted by the action of the group $\Sigma_3$. The three $G_2$ holonomy metrics are invariant under elements of order two $\sigma_{ij}$, which 
 are reflections about each of the three axis. The intersection point of the three branches corresponds to  the singular $G_2$ cone.
In the notation of \cite{Cvetic:2001ya}: $X_1$ corresponds to $\lambda =0$,
$X_2$ corresponds to $\lambda =-1$, and $X_3$ corresponds to  $\lambda =1$.}
\label{g2mod}
\end{center}
\end{figure}

An explicit $G_2$ holonomy  metric on the spin bundle over $S^3$ was
constructed in \cite{Bryand:1989mv,Gibbons:1989er}. 
In  \cite{Cvetic:2001ya} were presented  $G_2$ holonomy metrics on each $X_i$, characterised 
by three distinct values of a parameter $\lambda = 0, \pm 1$. We will re-derive those
results in a way that will be suitable for a generalisation 
to be discussed in the next section.
We define the following left-invariant $SU(2)$-valued one-forms\footnote{The one-form $\Sigma_3$ should not be confused with the triality group $\Sigma_3$.} on $SU(2)^3$
\be
a_1^{-1}da_1  \equiv  -\frac{i}{2} \sigma_i \tau_i~, ~~~~~~~~ 
a_2 da_2^{-1}  \equiv  -\frac{i}{2} \Sigma_i \tau_i~,~~~~~~~~ 
a_3^{-1}da_3  \equiv  -\frac{i}{2} \gamma_i \tau_i~,
\label{defoneforms}
\ee
where $\tau_i$ are the Pauli matrices. We can ``solve''  the constraint (\ref{constr}) by introducing 
two sets of angular variables parameterising the first two $SU(2)$ factors. Then more explicitly we have
\be
\begin{aligned}
\sigma_1 + i \sigma_2 & = e^{-i\psi_1} (d\theta_1+ i \sin\theta_1 d\phi_1)~,   ~~~~~~~~~\sigma_3 = d\psi_1 + \cos\theta_1 d\phi_1~,\\
\Sigma_1 + i \Sigma_2 & =  e^{-i\psi_2} (d\theta_2+ i \sin\theta_2 d\phi_2) ~,  ~~~~~~~~~\Sigma_3 =  d\psi_2 + \cos\theta_2 d\phi_2~,
\end{aligned}
\ee
obeying  $d\sigma_3 = - \sigma_1 \wedge \sigma_2, d\Sigma_3 = - \Sigma_1 \wedge \Sigma_2$ and cyclic permutations.  We also have $\gamma_i = M_{ij} (\Sigma_i - \sigma_i)$, 
where $M_{ij}$ is an $SO(3)$ matrix.
See  Appendix \ref{appdetails} for more details.
Introducing the notation 
\bea
da_1^2 \equiv 2 \sum_{i=1}^3 \sigma_i^2~, ~~~~~~~da_2^2 \equiv 2 \sum_{i=1}^3 \Sigma_i^2~,~~~~~~~da_3^2 \equiv 2 \sum_{i=1}^3 (\Sigma_i - \sigma_i)^2~,
\eea
we can write a metric ansatz  in a manifestly $\Sigma_3$ covariant form as
\bea
ds^2 = dt^2 + f_1 \, da_1^2 +f_2 \, da_2^2 + f_3 \, da_3^2 ~,
\eea
where $f_i(t)$ are three functions. In order to write the $G_2$ forms compatible with this metric, it is convenient to pass to a different set of functions $a,b,\omega$, defined by 
\bea
f_1 =  \frac{a^2}{2} + \frac{b^2}{8}(\omega^2-1)~,\qquad~~~~ f_2 = \frac{b^2}{4} (1-\omega) ~,\qquad ~~~~f_3 = \frac{b^2}{4} (1+\omega)~.
\eea
In terms of these the metric then reads
\bea \label{eq:Metricwitha}
ds^2  =  dt^2 + a^2\sum_{i=1}^3 \sigma_i^2  + b^2\sum_{i=1}^3 \left(\Sigma_i -\tfrac{1}{2}(1+\omega) \sigma_i\right)^2 
\eea
and we can introduce an orthonormal frame defined as
\bea
e^t  =  dt ~, \qquad~~~ \tilde e^a  =  a \sigma_a~, \qquad~~~  e^a =   b (\Sigma_a  -\tfrac{1}{2}(1+\omega) \sigma_a)~,
\label{sframe}
\eea
where $a$ is a tangent space index. 
The associative three-form $\phi$ may be conveniently written in terms of an auxiliary $SU(3)$ structure as  follows
\bea
\phi =  e^t \wedge J + \mathrm{Re} [e^{i\theta} \Omega]~.
\label{defg2str}
\eea
Here $\theta$ is a phase that needs not be constant and 
the differential forms $J,\Omega$ define the $SU(3)$ structure.
In terms of the local frame they read
\bea \label{eq:SU3forms}
J  =  \sum_{a=1}^3 e^a \wedge \tilde e^a~,~~~~ \qquad \Omega  =  (e^1 + i \tilde e^1 ) \wedge (e^2 + i \tilde e^2 ) \wedge (e^3 + i \tilde e^3 )~.
\eea
We can also rewrite 
\be
\begin{aligned}
\phi & =  e^t \wedge J + \cos\theta  \mathrm{Re} [\Omega] - \sin\theta \mathrm{Im} [\Omega]~,\\
* \phi & =  \frac{1}{2}J \wedge J + \left(\sin\theta  \mathrm{Re} [\Omega] + \cos\theta \mathrm{Im} [\Omega]\right)\wedge e^t~.
\end{aligned}
\ee
Imposing  $d\phi=d*\phi=0$ gives the following system of first order differential equations 
\be
\begin{aligned}
	Z  f_1' & = \frac{f_2 f_3}{\sqrt{2}(f_2 + f_3)}~,\\
        Z  f_2' & = \frac{f_3 f_1}{\sqrt{2}(f_3 + f_1)}~,\\
        Z  f_3' & = \frac{f_1 f_2}{\sqrt{2}(f_1 + f_2)}~, 
\end{aligned} 
\label{g2odes}
\ee
where the prime denotes derivative with respect to $t$ and 
\bea
Z \equiv \frac {f_1 f_2 + f_2 f_3 +  f_3 f_1}{\sqrt{(f_1 + f_2)(f_2+f_3)(f_3 + f_1)}}~.
\eea
This system can be integrated explicitly in terms of three constants
 $c_i$ defined as 
\bea
c_1^2 = \frac{(f_2-f_3)^2(f_2+f_3) }{(f_1+f_2)(f_1+f_3)} (f_1 f_2 + f_2 f_3 +  f_3 f_1) 
\eea
and $c_2,c_3$ obtained by cyclic permutations of this expression.  
Although a priori there are three independent integration constants, 
 at least one of them must vanish, and the other two are then equal. For example, assuming that $c_1=0$, then $f_2=f_3$ and 
\bea
\frac{r_0^6}{216} \equiv c_2^2 = c_3^2 = (f_1-f_2)^2 (f_1+\frac{f_2}{2}) ~.
\eea
After a change of radial coordinate the metric can be written in the form \cite{Atiyah:2001qf}
\bea
\label{eq:AtiyahWittenmetric}
	ds^2 (X_1) = \frac{dr^2}{1-(r_0/r)^3} + \frac{r^2}{72} \big( 1 -(r_0/r)^3 \big) \big( 2 da_2^2 - da_1^2 + 2 da_3^2 \big) + \frac{r^2}{24} da_1^2~.
\eea
This is a $G_2$ holonomy metric on the manifold $X_1$, where the three-sphere $C_1$ 
shrinks to zero at the origin and  the three-sphere
``at the centre'' is homologous to $C_2$ or $-C_3$. In the notation of 
\cite{Cvetic:2001ya} this corresponds to the metric with $\lambda=0$. This metric is invariant 
under $\sigma_{23}$, or equivalently, under the interchange of $f_2\leftrightarrow f_3$.  In fact, we 
can re-define the triality symmetry $\Sigma_3$ as acting on the functions $f_1,f_2, f_3$
in the obvious way\footnote{For the elements of order two, we must also reverse the orientation 
of the seven-dimensional space.}. 
The other two solutions may be obtained for example  by acting with the cyclic permutation $\sigma_{231}$. 
In the notation of \cite{Cvetic:2001ya} 
the metric on $X_2$ corresponds to $\lambda =-1$ and  
the metric on $X_3$ to  $\lambda =1$.  Notice that the phase $\theta$ that enters in the definition of the $G_2$ structure in (\ref{defg2str})
is not constant, and in particular we have 
\bea	
\cos \theta = \lambda \frac{3\sqrt{3}r_0^3 r^{3/2}}{(4r^3-r_0^3)^{3/2}} 
\eea
with $\lambda=0,\pm 1$, respectively. 
The conical metric
\bea
\label{conemetric}
	ds^2_\mathrm{cone} = dr^2 + \frac{r^2}{36} \big( da_1^2 + da_2^2 + da_3^2 \big) 
\eea
 may be obtained from any of these by setting $r_0=0$ and is invariant under $\Sigma_3$.  
The parameter space of these metrics is depicted in Figure \ref{g2mod}. 

In the figure, moving along an axis corresponds to changing the radius of the three-sphere at the centre of one of the $X_i$ spaces, 
whose volume is $\tfrac{2}{3\sqrt{3}}\pi^2 r_0^3$. Hence, in analogy with the deformed conifold,
$r_0$ measures the amount of ``deformation'' of the conical singularity.  However, in analogy with the resolved conifold, we can also 
define a parameter measuring the amount of ``resolution'' of a  space $X_i$. 
Recall that in the resolved conifold 
the resolution parameter may be defined as the difference of volumes of two two-spheres at large distances 
\cite{Maldacena:2009mw}. In particular, this 
parameter measures the breaking of a $\Z_2$ symmetry of the singular (and deformed) conifold, consisting in swapping these two-spheres. 
Here we can define a triplet of resolution parameters, each measuring the breaking of the $\Z_2$ symmetry given by reflection about the three axis in  
Figure \ref{g2mod}. Following \cite{Atiyah:2001qf}, we first consider the 
``volume defects'' of the  sub-manifolds $C_i^\infty$ defined at a large constant value of $r$. In terms of the 
radial coordinate $t$ we have the following asymptotic form of the $G_2$ metrics
\bea
ds^2(X_i) = dt^2 + \frac{t^2}{36}\left[ da_1^2 + da_2^2 + da_3^2  - \frac{r_0^3}{2t^3}  \left( \ell_1 da_1^2 + \ell_2 da_2^2 + \ell_3 da_3^2 \right) + O(r_0^6/t^6)\right]
\eea
where $t = r - r_0^3/(4r^2)+ O(r^{-5})$
and the constants $(\ell_1, \ell_2,\ell_3)$ take the values: $(-2,1,1)$ for $X_1$, $(1,-2,1)$ for $X_2$, and  $(1,1,-2)$ for $X_3$. 
Then for the ``volume  defects'' we have
\bea
\vol(C_i^\infty) = \frac{16}{27}\pi^2 t^3 + \frac{2}{9}\pi^2 r_0^3 \ell_i
\eea
and we may define the $i$th resolution parameter as
\bea
\alpha^\mathrm{res}_i \equiv   \vol(C_{i+1}^\infty) - \vol(C_{i-1}^\infty) = \frac{2}{9}\pi^2 r_0^3 (\ell_{i+1} - \ell_{i-1})~.
\eea
To see that this is a sensible definition, let us evaluate $\alpha^\mathrm{res}_1$ in the three cases $X_i$. We have
\bea
\alpha^\mathrm{res}_1 (X_1) = 0 ~, \qquad \alpha^\mathrm{res}_1 (X_2) = - \frac{2}{3} \pi^2 r_0^3 ~, \qquad \alpha^\mathrm{res}_1 (X_3) = + \frac{2}{3} \pi^2 r_0^3 ~.
\eea
The interpretation is that the manifold $X_1$ preserves the $\Z_2$ reflection about the axis 1, hence from this point of view, $r_0$ is a ``deformation'' parameter. On the other hand, the manifolds $X_2$ and $X_3$ break this symmetry in opposite directions. From the point of view of $X_1$, $X_2$ is
 a ``resolution'' and $X_3$ its flopped version.  Notice in particular that we cannot have ``resolution'' and ``deformation'' at the same time, exactly as 
it happens for the conifold in six dimensions. Indeed, the relation to the conifold may be made very precise by considering the different $G_2$ holonomy metrics 
(times $\R^{1,3}$) as solutions of M-theory. Then there exist three different reductions to Type IIA that give rise to manifolds with topologies of the deformed, resolved, and flopped resolved conifold, respectively  \cite{Atiyah:2001qf,Atiyah:2000zz}.

Finally, let us recall some facts about the cohomologies of these spaces. 
For each $X_i$ the third cohomology group is $H^3(X_i;\Z)=\Z$, so there is only one
generator, that can be chosen to integrate to one on the non-trivial three-cycle. 
However, it is convenient to introduce the following set of three-forms 
\be
\begin{aligned}
\eta^1 = -\frac{1}{16\pi^2}  \sigma_1 \wedge \sigma_2 \wedge \sigma_3~, \quad~ \eta^2  = \frac{1}{16\pi^2}  \Sigma_1 \wedge \Sigma_2 \wedge \Sigma_3~,
\quad ~ \eta^3  = -\frac{1}{16\pi^2}  \gamma_1 \wedge \gamma_2 \wedge \gamma_3 ~, 
\end{aligned}
\ee
which are exchanged by the action of $\Sigma_3$. We also have that 
\bea
\eta^3  =  -\frac{1}{16\pi^2}  (\Sigma_1- \sigma_1 )\wedge (\Sigma_2-\sigma_2 )\wedge (\Sigma_3- \sigma_3) ~.
\eea
Integrating the $\eta^j$ over the sub-manifolds $C_i,$ we have the relation
\be \label{eq:integral}
	\int_{C_i} \eta^j = \delta_{j,i+1} - \delta_{j,i-1}
\ee
and by Poincar\'e duality $\eta^i\to C_i$ the intersection numbers $C_i \cdot C_j = \delta_{j,i+1} - \delta_{j,i-1}$. Notice that having fixed the orientation 
so that  $C_1 \cdot C_2 =+1$, we then have that $C_3 \cdot C_1 = +1$, which gives $\int_{C_3}\sigma_1 \wedge \sigma_2 \wedge \sigma_3 = -16\pi^2$
so that the orientation of $C_3$ is opposite to that of $C_2$.

\section{Fivebranes wrapped on a three-sphere in $S^3\times \R^4$}
\label{solutions:sec}

In this section we will discuss solutions describing
fivebranes wrapped on a three-sphere inside a  $G_2$ holonomy manifold $X_i$.
The backreaction of the fivebranes modifies the geometry, making the internal space a smooth 
\emph{torsional $G_2$ manifold}. 
As we will see, the topology is again $\R^4\times S^3$, although we will be careful about 
which $S^3$. Since we are interested in solutions arising from NS fivebranes, we 
 may work in Type I supergravity, 
and allow for a non-trivial three-form $H_3$ and dilaton profile. 
By applying an S-duality to the NS5 branes in Type IIB, these solutions may also be interseted as arising 
from D5 branes.

\subsection{Torsional $G_2$ solutions}

General classes of supersymmetric solutions of Type I and heterotic supergravities have been studied in \cite{Gauntlett:2002sc}, extending the 
works of \cite{Strominger:1986uh} and \cite{Hull:1986kz}. Here we are interested in solutions where the non-trivial geometry is seven-dimensional
and is characterised by a $G_2$ structure. We  will therefore refer to this class as \emph{torsional $G_2$ solutions}. 
The ten-dimensional metric in string frame is unwarped
\be
ds^2_{str}  =  dx^2_{1+2} + ds^2_7~.
\ee
The supersymmetry equations are equivalent to exterior differential equations obeyed by the 
 $G_2$ structure on the seven-dimensional space with metric $ds^2_7$ and read
\cite{Gauntlett:2002sc,Friedrich:2001yp}:
\be
\begin{aligned}
\phi \wedge d\phi   & =  0 \\
d (e^{-2\Phi} *_7 \phi) & =  0 \\
e^{2\Phi} *_7 d(e^{-2\Phi} \phi)& =  -  H_3 
\end{aligned}
\label{torsional}
\ee
where $\Phi$ is the dilaton field and $*_7$ denotes the Hodge star operator with respect to the metric $ds^2_7$.
 The Bianchi identity $dH_3=0$ implies 
that all remaining equations of motion are satisfied. See \cite{Gauntlett:2002sc} for a more detailed discussion of this type of $G_2$ structure.  
Examples of solutions to these equations were discussed in \cite{Acharya:2000mu,Maldacena:2001pb,Gauntlett:2002sc,Canoura:2008at} and we
shall return to some of these later.

\subsection{Ansatz and BPS equations}
\label{bpsses}

We will now specify ansatze for the metric, $G_2$ structure and $H_3$ field. 
Although the $G_2$ structure determines the metric uniquely, and the $H_3$ field is then derived from the third equation in 
(\ref{torsional}), we find more convenient to start with an ansatz for $H_3$ that is manifestly closed $dH_3=0$.
Specifically, we use the ansatz for the metric and associative three-form
discussed earlier 
\be
\label{metansatz}
\begin{aligned}
ds^2_7  & =  M \big[ dt^2 + f_1 \, da_1^2 +f_2 \, da_2^2 + f_3 \, da_3^2 \big]\\
\phi  &  =   M^{3/2} \left[e^t \wedge J + \mathrm{Re} [e^{i\theta} \Omega]\right]
\end{aligned}
\ee
where we inserted a 
factor of $M$ in front of the metric. For the  three-form flux we take
\be
	H_3 = 2 \pi^2 M  \bigg[  \gamma_1 \, \eta^1 +  \gamma_2 \, \eta^2 +\gamma_3 \, \eta^3  + \gamma_4 dt \wedge 
	\sum_{i=1}^3 \sigma_i \wedge \Sigma_i  \bigg]\label{fluxansatz}
\ee
where the factor of $2\pi^2M$ is again for convenience.  Imposing $dH_3=0$ implies 
\be 
\gamma_i = \gamma + \alpha_i\qquad i=1,2,3~, \qquad ~~~~ \gamma_4 =  \frac{\gamma'}{16 \pi^2}~,
\ee
where $\gamma$ is a function and $\alpha_i$ are three integration constants. The ansatz then depends on four functions 
$f_i,\gamma$ and three constants $\alpha_i$, although we will see
 that the homology relation among the $C_i$ implies that  only two of these are significant,
and are  fixed by flux quantisation and regularity of the metric.

The action of the $\Sigma_3$ symmetry  on the functions in the ansatz is given by
\be
\begin{aligned}
\sigma_{12} : & \,   (f_1,f_2,f_3) \, \rightarrow \, (f_2,f_1,f_3) \qquad\qquad~  &\sigma_{231}  :  (f_1,f_2,f_3) \, \rightarrow \, (f_2,f_3,f_1) \\
\sigma_{12} : & \,   (\gamma_1,\gamma_2,\gamma_3) \, \rightarrow \, (-\gamma_2,-\gamma_1,-\gamma_3) \qquad~~
  & \;\;\;\sigma_{231}  \, : (\gamma_1,\gamma_2,\gamma_3)  \, \rightarrow \,   (\gamma_2,\gamma_3,\gamma_1) 
\end{aligned}
\ee
with the rest following from group multiplication rules. 
The minus signs in the action of the order two elements on the $\gamma_i$ arise because the orientation of the seven-dimensional space is reversed,
 hence  the Hodge $*_7$ operator in (\ref{torsional}) changes sign.

Inserting the ansatz into the equations (\ref{torsional}), after some computations, we arrive at a system of first order ODEs. 
We relegated some details in Appendix \ref{tedious}.
We have four  coupled ODEs for the  functions $f_i,\gamma$, while an additional
 decoupled equation determines the dilaton profile in terms of the other 
functions. Although the explicit form of the equations is rather complicated, 
its presentation can be simplified slightly by organising it in terms of the $\Sigma_3$ symmetry. We have 
\be
\begin{aligned}
\label{eq:BPSwithf}
	D(f_i,  \gamma_i) \, f_1'    &  =  F(f_1,f_2,f_3, \gamma_1, \gamma_2,\gamma_3) \\
	D(f_i,  \gamma_i) \, f_2'    &  =  F(f_2,f_1,f_3, -\gamma_2, -\gamma_1,-\gamma_3) \\
	D(f_i,  \gamma_i) \, f_3'    &  =  F(f_3,f_2,f_1, -\gamma_3, -\gamma_2,-\gamma_1)\\
	D(f_i,  \gamma_i) \, \gamma' & =   G (f_i,  \gamma_i) 
	\end{aligned}
\ee
where, defining  $Q\equiv f_1 f_2 + f_2 f_3 + f_3 f_1$, the functions   $F$, $G$ and  $D$ are given by 
\be
	\begin{aligned}
		F(f_1,f_2,f_3, \gamma_1, \gamma_2,\gamma_3) &=\, 768 f_2 f_3 (f_1+f_2)(f_1+f_3) + \gamma_1^2 (f_2+f_3)^2 + \gamma_2 \gamma_3 (3 f_1^2-Q) \\
		&+ \gamma_1 \gamma_2 (Q + 2 f_1 f_3 -f_3^2) + \gamma_1 \gamma_3 (Q + 2 f_1 f_2 -f_2^2) \\
		&+ 32 \gamma_1 Q (f_3-f_2) - 32 \gamma_2 Q (f_1+f_3) + 32 \gamma_3 Q (f_1+f_2) ~,\\[2mm]
		G(f_i, \gamma_i)  &= \,-256 \big[ \gamma_1 (f_2+f_3) (Q+f_2f_3) + \gamma_2 (f_3+f_1) (Q+f_3f_1)~, \\
\end{aligned}
\ee
and 
\be
	\begin{aligned}
		\sqrt{2} D(f_i, \gamma_i) &=\, 32 \big( \gamma_1^2 (f_2 + f_3)^3 + \gamma_2^2 (f_3 + f_1)^3 + \gamma_3^2 (f_1 + f_2)^3 \\
		&+ 2 \gamma_1 \gamma_2 f_3 (3 Q - f_3^2) + 2 \gamma_1 \gamma_3 f_2 (3 Q - f_2^2) + 2 \gamma_2 \gamma_3 f_1 (3 Q - f_1^2) \\
		&+ 96 Q \left(\gamma_1 (f_3^2 - f_2^2) + \gamma_2 (f_1^2 - f_3^2) + \gamma_3 (f_2^2 - f_1^2) \right) \\
		&+ 2304 Q (f_1+f_2)(f_2+f_3)(f_3+f_1) \big)^{1/2} Q^{1/2}~.
	\end{aligned}
\ee
The decoupled equation for the dilaton reads
\be
	2 \sqrt{2} \, Q D(f_i,  \gamma_i) \, \Phi' =  P(f_i,\gamma_i)\label{dilet}
\ee
where 
\be
\begin{aligned}
P(f_i, \gamma_i) & = \, 2 \gamma_1^2 (f_2 + f_3)^3 +2 \gamma_2^2 (f_3 + f_1)^3+2 \gamma_3^2 (f_1 + f_2)^3 \\
		&+4 \gamma_1 \gamma_2 f_3 (3Q-f_3^2) +4 \gamma_2 \gamma_3 f_1 (3Q-f_1^2) +4 \gamma_3 \gamma_1 f_2 (3Q-f_2^2) \\
		&+ 96 Q \left(\gamma_1 (f_3^2 - f_2^2) + \gamma_2 (f_1^2 - f_3^2) + \gamma_3 (f_2^2 - f_1^2) \right) ~.
\end{aligned}
\ee
Once a solution for $f_i,\gamma$ is determined (for example numerically), then the dilaton can be obtained integrating \eqref{dilet}. 
Notice that $D$ and $P$ are invariant under $\Sigma_3$ and $G$ is invariant up to an overall change of sign under transformations
of order two elements. It  follows that $\Phi$ is invariant under $\Sigma_3$. The phase  $\theta$ in the associative three-form is a non-trivial  function of $f_i,\gamma_i$, whose explicit form can be found in Appendix \ref{tedious}. 

From the BPS system it is clear that generically, for any given solution we have in fact \emph{six} different solutions, 
obtained acting with $\Sigma_3$. To study the system we can therefore focus on one particular case. 
Notice that if we formally set $\gamma_i=0$ in (\ref{eq:BPSwithf}), then we recover the $G_2$ holonomy BPS equations (\ref{g2odes}). 
Solutions to this system were presented in \cite{Acharya:2000mu}, \cite{Maldacena:2001pb,Chamseddine:2001hk} 
and \cite{Canoura:2008at}. In particular, the solution of \cite{Maldacena:2001pb} corresponds to the near brane limit of
 a configuration  of $M$ NS fivebranes wrapped on an $S^3$ inside the $G_2$ manifold $S^3\times \R^4$. 
Below we will be  more precise about which  $G_2$ manifold $X_i$ is relevant
 for a particular solution of the type discussed in \cite{Maldacena:2001pb}.

\subsubsection*{Maldacena-Nastase solutions}

The basic solution of  \cite{Maldacena:2001pb} may be recovered from our general ansatz by setting
\be
	\begin{aligned}
		f_2 + f_3 &= 1/8 ~,\\
		\alpha_1 = \alpha_2 &= - \alpha_3 = -1~.
	\end{aligned}
\ee
For consistency these conditions impose also $\gamma_2 = \gamma -1= - 16 f_2$. 
Then we are left with two unknown functions, $f_1$ and $\gamma$. To make contact
with the variables in  \cite{Maldacena:2001pb} one has to set 
\be
	\begin{aligned}
		f_1 &= \frac{4 R^2 + \gamma^2 - 1}{32} 
	\end{aligned}
\ee
and then passing to the variables $a,b,\omega$ we have
\bea
a^2 = \frac{R^2}{4} ~, ~~~\qquad b^2 = \frac{1}{4} ~, ~~~\qquad \omega = \gamma~,
\eea
so that the metric reduces to the ansatz\footnote{Our one-forms are related to those in \cite{Maldacena:2001pb} as:  $\sigma_{i}^\mathrm{here}= w_i^\mathrm{there}$ and $\Sigma_i^\mathrm{here}=\tilde w_i^\mathrm{there}$.}  in \cite{Maldacena:2001pb}, namely 
\bea
ds^2_7 = M \left[ dt^2 + \frac{R^2}{4}\sum_{i=1}^3 \sigma_i^2 + \frac{1}{4}\sum_{i=1}^3 \left(\Sigma_i - \tfrac{1}{2}(1+\omega)\sigma_i\right)^2 \right]~.
\eea
Similarly, the $H_3$ reduces to that in  \cite{Maldacena:2001pb}.
In terms of the functions $R$ and $\omega$, the system \eqref{eq:BPSwithf} reduces to the system in the 
Appendix of \cite{Maldacena:2001pb}.
As discussed in \cite{Maldacena:2001pb,Chamseddine:2001hk} there exists a unique non-singular 
solution to the differential equations. 
In the interior\footnote{Asymptotically the geometry is a linear dilaton background.}
the topology of the solution is $S^3\times \R^4$, where the three-sphere\footnote{Notice that as $t\to 0$ we have: $f_1 \to 0$, $f_2\to 0$ and $f_3\to 1/8$ \cite{Maldacena:2001pb,Chamseddine:2001hk}.} $C_3$ smoothly shrinks to zero  and the three-cycle is represented by $C_1$ or $-C_2$. Then, more precisely, the topology of this solution is that of the $G_2$ manifold $X_3$. 
 The authors of \cite{Maldacena:2001pb} discussed also a second solution  
which  can be obtained from the basic solution by acting with $\sigma_{23}$. Hence 
the topology of this solution is that of $G_2$ manifold $X_2$. 
It is clear that there are four more different solutions, obtained by acting with elements of $\Sigma_3$.

\subsection{One-parameter families of solutions}
\label{sec:Expansions}

Finding analytic solutions to the BPS system \eqref{eq:BPSwithf} seems very difficult. As usual in these cases, we will then turn to a combination of numerical methods and asymptotic expansions. We are interested in non-singular solutions to the system, giving 
rise to spaces with topology 
$S^3\times \R^4$. As for the  $G_2$ holonomy manifolds $X_i$ and the 
Maldacena-Nastase solutions, we then require that two functions $f_i$
go to zero in the interior, while the third function approaches a constant value, parameterising the size of the non-trivial 
$S^3$ inside $S^3\times \R^4$. We can restrict our attention to one particular case, for example we may 
require that $f_1$ and $f_2$ go  to zero in the IR (at $t=0$) while $f_3$ 
approaches a constant value $f_3(0) \equiv  c>0$. 
This then has the topology of $X_3$, where $C_3$ shrinks to zero. 
This solution was studied in \cite{Canoura:2008at}.

More generally, we impose boundary conditions such that the topology of the solution is that of one of the manifolds $X_i$.   This fixes the values of the integration constants $\alpha_i$.
Using the relation (\ref{eq:integral})  we can 
evaluate the flux of the three-form $H_3$ (\ref{fluxansatz}) on the sub-manifolds $C_i$, defined  exactly like in  \eqref{subma},
 and at some constant $t$. We have 
\be
		q_i \equiv \frac{1}{4 \pi^2} \int_{C_i} H_3 =  \frac{M} {2} (\gamma_{i+1} - \gamma_{i-1}) =  \frac{M} {2} (\alpha_{i+1} - \alpha_{i-1}) 
	\ee
where the result does not depend on $t$. The $q_i$ then obey the relation $q_1+q_2+q_3=0$, reflecting
 the homology relation $[C_1]+ [C_2]+ [C_3]=0$. Hence 
we can parameterise the constants $\alpha_i$ by taking for example
\bea
\alpha_1 = - k_1~, \qquad \alpha_2 = - k_2~, \qquad \alpha_3 = k_2~,
\label{defalfa}
\eea
so that
\bea
q_1 = - Mk_2~,~~   \qquad q_2 = \frac{M}{2}( k_1 +k_2)~,~~        \qquad q_3 = \frac{M}{2}( -k_1 +k_2)~.
\eea
The constants $k_1,k_2$ are determined for a given solution as follows.  Suppose that 
we require the manifold to have the topology of $X_3$.  Then the flux of $H_3$ through $C_1$ is
 minus the flux through $C_2$, namely  $q_1=-q_2$. 
In terms of the constants $k_1,k_2$, then we must have
$k_1 = k_2 = k$ and $k$   can be 
 reabsorbed in the definition of $M$.
There are then essentially two choices for $k$, namely
$k=\pm 1$, corresponding to two different solutions, both with topology of $X_3$. We denote these two solutions as $X_{31}$ and $X_{32}$, respectively. 
More generally, there are six different solutions and  we  denote the corresponding spaces as $X_{ij}$. 
The topology of the spaces $X_{ij}$ is the same as the $G_2$ holonomy manifolds $X_i$. 

In each case the flux through the non-trivial cycle must be quantised, thus we require that 
\bea
{\cal N}(X_{ij}) = |\epsilon_{ijk}|\frac{1}{4\pi^2} \int_{C_k} H_3 = M~.
\eea
The signs have been chosen to give always a positive number and are consistent with the 
action of $\Sigma_3$.  
In conclusion, flux quantisation, together with the condition that the flux through the vanishing three-sphere vanishes, fixes the integration constants $k_1, k_2$ 
in all cases. We summarise the values of the $k_1,k_2$ and the  $q_i$ 
in each of the six solutions in Table \ref{kvalues}.

\begin{table}[ht!]
 \begin{center}
	\begin{tabular}{|c|c|c|c|c|c|}
		\hline
			& $ k_1$ & $k_2$ & $q_1$ & $q_2$ & $q_3$ \\
		\hline
			$X_{31}$ & $1$ & $1$ & $-M$ & $M$ & $0$ \\
		\hline
			 $X_{21}$   & $-1$ & $1$ & $-M$ & $0$ & $M$ \\
		\hline	
			$X_{12}$ & $-2$ & $0$ & $0$ & $-M$ & $M$ \\
			\hline
			$X_{32}$  & $-1$ & $-1$ & $M$ & $-M$ & $0$ \\
			\hline
		        $X_{23}$  & $1$ & $-1$ & $M$ & $0$ & $-M$ \\
		\hline		
			$X_{13}$  & $2$ & $0$ & $0$ & $M$ & $-M$ \\
		\hline
	\end{tabular}
\end{center}
\caption{Values of $k_1,k_2$ and $q_i$ for the six different solutions $X_{ij}$. The basic Maldacena-Nastase solution is the $c=1/8$ limit of 
$X_{31}$,  while the second Maldacena-Nastase solution is the $c=1/8$ limit of 
$X_{21}$.}
\label{kvalues}
\end{table}

\subsubsection{Expansions in the IR}

For definiteness, let us concentrate on the case of $X_{31}$. 
To discuss the expansions it is convenient to first rescale 
the radial coordinate by a constant factor as  $t \rightarrow \sqrt{c} t$.
The boundary conditions that we impose  at $t=0$ determine the
expansions\footnote{If we had left the constants $k_1, k_2$ arbitrary, the IR expansions of the functions $f_i, \gamma$ in power
series, would impose $\gamma_{(0)} = k_1=  k_2$,   that is $q_3 = 0$.}
  of the functions $f_i$ and $\gamma$ around $t=0$ as follows 
\be
	\begin{aligned}
		f_1 + f_2 &= \frac{1}{8} c t^2 + \frac{1 - 384 c^2}{147456 c} t^4 + O \left( t^{6} \right) \\
		f_1 - f_2 &= \frac{1}{96} t^2 + \frac{3 -256 c^2}{589824 c^2} t^4 + O \left( t^{6} \right) \\
		f_3 &= c + \frac{-5 + 192 c^2}{6144c} t^2 + \frac{-3 + 224 c^2 +2048 c^4}{6291456 c^3} t^4 +O \left( t^{6} \right) \\
		\gamma &= 1 - \frac{1}{24} t^2 + \frac{-1+ 128 c^2}{49152 c^2} t^4 + O \left( t^{6} \right)
	\end{aligned}
\label{irexp}
\ee
%
%
The corresponding expansion for the dilaton reads
\be
	\Phi = \Phi_0  - \frac{7}{12288 c^2} t^2 + \frac{-293 + 21504 c^2}{452984832 c^4} t^4 + O \left( t^{6} \right)
\ee
%
where $\Phi_0$ is an (IR) integration constant. We therefore have a family of non-singular solutions, 
parameterised by the constant $c$, measuring the size of the non-trivial $S^3$. 
Using numerical methods one can then check that, for any value of $c \geq 1/8$, 
there exists a non-singular solution approaching \eqref{irexp} as $t\to 0$.
The special value $c=1/8$  corresponds precisely to the Maldacena-Nastase solution. 
Hence we have a one-parameter family of solutions with topology of $X_3$ (in the interior),
generalising  the solution discussed in \cite{Maldacena:2001pb}.

\subsubsection{Expansions in the UV}

Towards infinity we find two different types of asymptotic expansions of the functions. 
In one expansion the functions have the following  behaviour for large $t$: 
\be
	\begin{aligned}
		f_1 &= \frac{c t^2}{36} + \frac{1}{4} - \frac{21}{16 c t^2} + O \left( t^{-4} \right) \\
		f_2 &= \frac{c t^2}{36} - \frac{1}{4} - \frac{21}{16 c t^2} + O \left( t^{-4} \right) \\
		f_3 &= \frac{c t^2}{36} + \frac{69}{16 c t^2} + O \left( t^{-4} \right) \\
		\gamma &= \frac{1}{3} + O \left( t^{-4} \right) \\
		\Phi &= \Phi_{\infty} + O \left( t^{-4} \right)
	\end{aligned}
	\label{genuv}
\ee
where $\Phi_{\infty}$ is an (UV) integration constant. 
 Notice that  the constant $c$ appears here trivially
 because of the rescaling $t \rightarrow \sqrt{c} t$ we made, therefore at this order we don't 
 see a genuine UV
 integration constant.  After this particular order the expansion in inverse powers of $t$ is not valid anymore and one would need to use other types of series to gain more precision.  This expansion can be matched numerically to the IR expansions for all values of $c>1/8$.

Already from these few orders we can extract some useful information. 
The functions $f_i$ all have the same leading behaviour in $t^2$ towards infinity, 
 corresponding  to the  $G_2$ holonomy cone.  
From the sub-leading terms we can also read off an  effective ``resolution parameter'', measuring the amount of $\Z_2$ symmetry breaking in each case.
The asymptotic form of the metric here is
\be
ds^2(X_{ij})_7 = Mc \left[ dt^2 + \frac{t^2}{36}(da_1^2 + da_2^2 + da_3^2 ) + \frac{1}{4c} \left( \ell_1 da_1^2 + \ell_2 da_2^2 + \ell_3 da_3^2 \right)  + O(1/t)\right]  
\ee 
where $(\ell_1, \ell_2,\ell_3)=(1,-1,0)$ for $X_{31}$ and the remaining ones are 
determined by the $\Sigma_3$ action. The ``volume  defects''  are given by
\bea
\vol(C_i^\infty) = (Mc)^{3/2}\left( \frac{16}{27}\pi^2 t^3 - 4\pi^2  \ell_i \frac{t}{c}\right)~.
\eea
Notice that even after subtracting the leading divergent part these volumes are now ``running''. This running is analogous to the running 
volume of the two-sphere at infinity in the resolved deformed conifold, although here the running is linear, rather than logarithmic.
Then the $i$th effective resolution parameter may be defined as
\bea
\alpha^\mathrm{res}_i \equiv   \vol(C_{i+1}^\infty) - \vol(C_{i-1}^\infty) = (Mc)^{3/2} 4 \pi^2  (\ell_{i-1} - \ell_{i+1}) \frac{t}{c}~.
\eea
In Section \ref{g2metrics:section} we saw that in the $G_2$ holonomy manifold $X_i$ the resolution parameter 
$\alpha^\mathrm{res}_i$ was vanishing, reflecting a $\Z_2\subset \Sigma_3$ symmetry of the geometry.
Hence, the  relevant\footnote{Of course $\alpha^\mathrm{res}_{i+1}$ and $\alpha^\mathrm{res}_{i-1}$ are also non-zero, since the solutions do not preserve
any $\Z_2$ symmetry. However these are less interesting parameters, since they are non-zero 
also in the $G_2$ holonomy manifold $X_i$.} 
resolution  parameter to consider for the manifolds $X_{ij}$ is $\alpha^\mathrm{res}_i$. For example, we find that
\bea
\alpha^\mathrm{res}_3  (X_{31}) = - \alpha^\mathrm{res}_3  (X_{32}) = 8 \pi^2  (Mc)^{3/2}\frac{t}{c}~,
\eea 
which is again running. Notice that keeping $Mc$ fixed, a non-zero value of 
the parameter $c^{-1}$ may then be interpreted as turning on a ``resolution'' in the manifold
$X_3$. Indeed, we will show below that the limit $c\to \infty$ gives the $G_2$ holonomy manifold $X_3$.

We also find a second type of expansion at large $t$, in which
the behaviour of the functions is different and we have
\be
	\begin{aligned}
		f_1 &= \frac{\sqrt{2}}{16} t + \kappa - \frac{\sqrt{2}}{32 t} + \frac{1-16\kappa}{32 t^2} + \frac{7-64\kappa + 512 \kappa^2}{64 \sqrt{2} t^3} + O \left( t^{-4} \right) \\
		\gamma &= \frac{\sqrt{2}}{2 t} + \frac{1-8\kappa}{t^2} + \frac{9 -64\kappa +256 \kappa^2}{2 \sqrt{2} t^3} + O \left( t^{-4} \right) \\
		\Phi &= -\frac{\sqrt{2}}{4} t + \frac{3}{8} \log t + \Phi_{\infty} + \frac{3(1+32 \kappa)}{16 \sqrt{2} t} + \frac{29 - 192 \kappa - 3072 \kappa^2}{128 t^2} \\
		&\quad + \frac{9 - 928 \kappa + 3072 \kappa^2 + 32768 \kappa^3}{128 \sqrt{2} t^3} + O \left( t^{-4} \right)
	\end{aligned}
	\label{uvs}
\ee
In this case the  expansions remain valid at high orders, hence presumably they don't break down.  
These expansions may be matched numerically to the IR expansions for the particular 
case $c=1/8$, thus they correspond to the Maldacena-Nastase 
solution\footnote{The factor of $\sqrt{8}$ difference with respect to the UV behaviour of the functions  
in \cite{Maldacena:2001pb} is due to the rescaling $t\to \sqrt{c}t$.}. 
Despite the fact that $\kappa$ seems a free constant, it can be determined  numerically to be $\kappa \approx -0.2189$.

\subsubsection{Numerical solutions}

As can be seen from the expansions, while the behaviour of the functions in the 
IR changes smoothly as we vary the parameter $c$,  the behaviour in the UV changes discontinuously if we 
choose the extremal\footnote{This behaviour is analogous to that 
of the one-parameter family of solutions discussed in \cite{Butti:2004pk,Maldacena:2009mw}. 
In that case, for the special value of the integration constant $\gamma^2=1$, one obtains 
the solution of \cite{Maldacena:2000yy}, which has linear dilaton asymptotics.} value for the parameter $c=1/8$. 
\begin{figure}[t!]
	\begin{tabular}{c c}
		\includegraphics[width=0.45\textwidth]{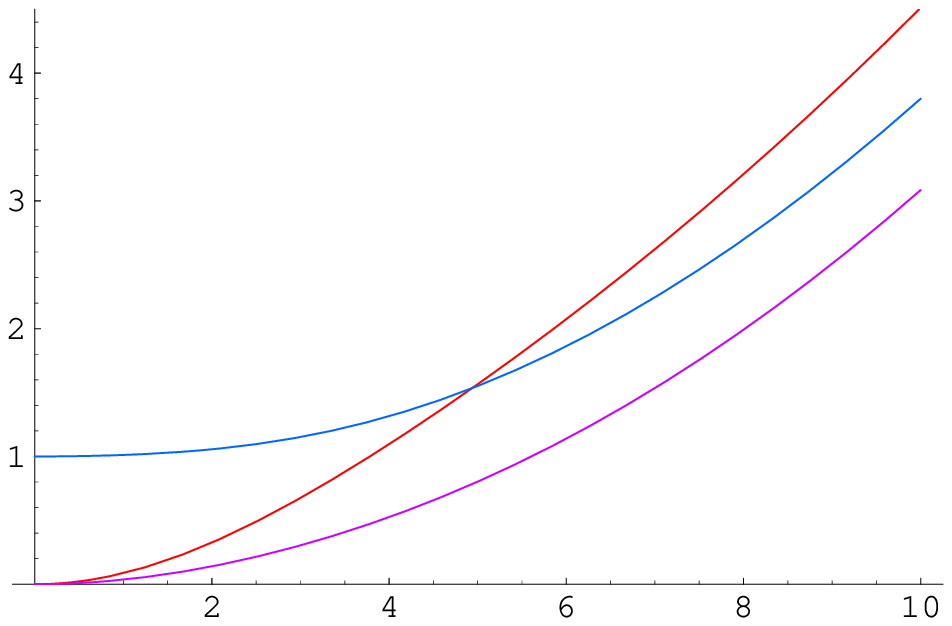} & \includegraphics[width=0.45\textwidth]{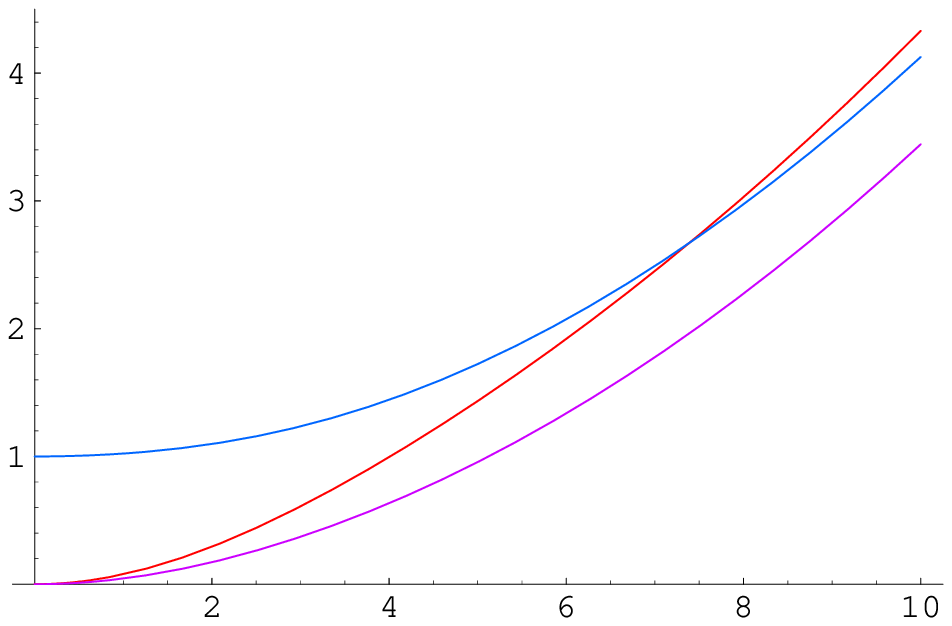} \\
		\includegraphics[width=0.45\textwidth]{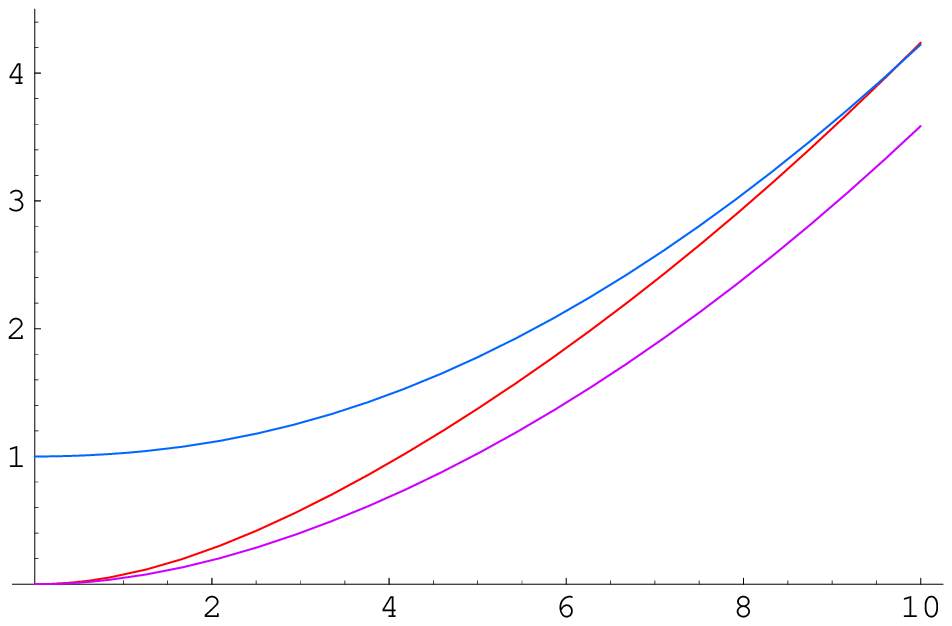} & \includegraphics[width=0.45\textwidth]{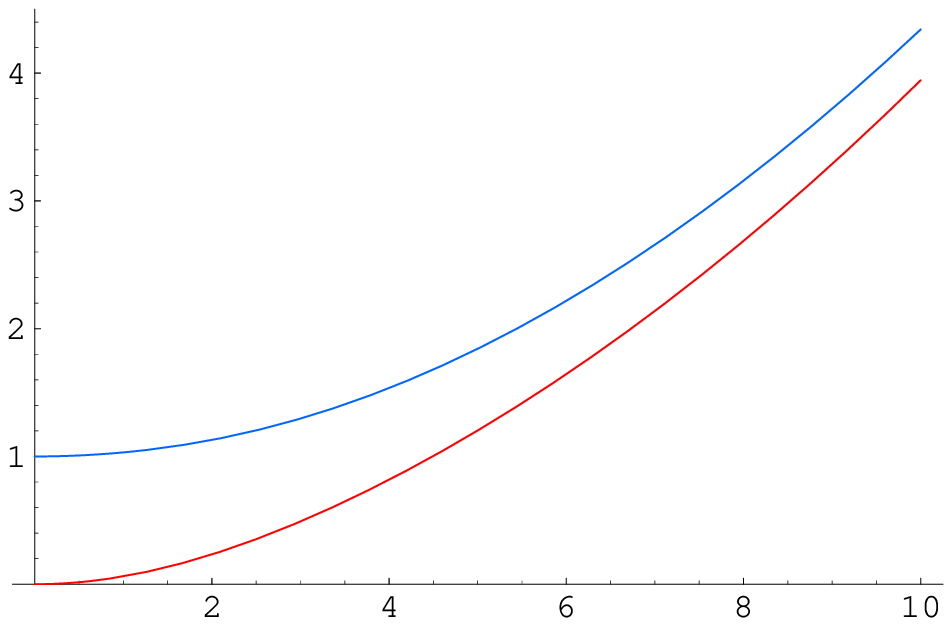}
	\end{tabular}
\caption{Plots of the functions $c^{-1} f_i$  for the $X_{31}$ solution for different values of $c$. $f_1$ is in red, $f_2$ in purple and $f_3$ in blue. The factor $c^{-1}$ is there for normalisation purposes. From the top left to the bottom right, the values of $c$ are increasing and are 0.2, 0.3, 0.4 for the first three. The bottom-right plot corresponds to the space $X_3$ where $f_1 = f_2$. This is formally the plot for $c \rightarrow \infty$.}
\label{fig:Largec}
\end{figure}
Here we present some plots of the numerical solutions 
to illustrate the qualitative behaviour of the metric functions $f_i$ for various values of $c$.

\begin{figure}[ht]
	\begin{tabular}{c c}
		\includegraphics[width=0.45\textwidth]{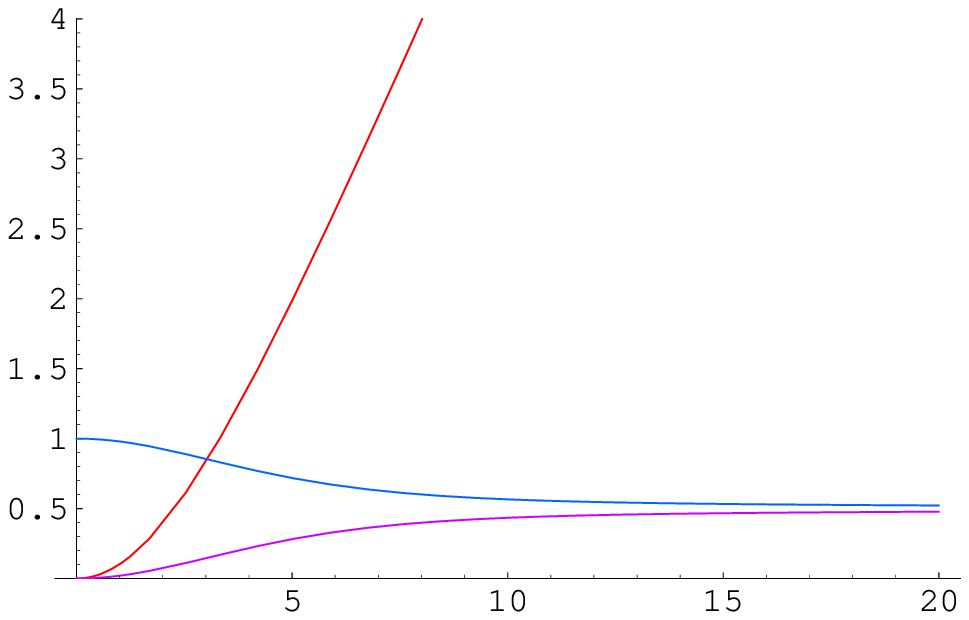} & \includegraphics[width=0.45\textwidth]{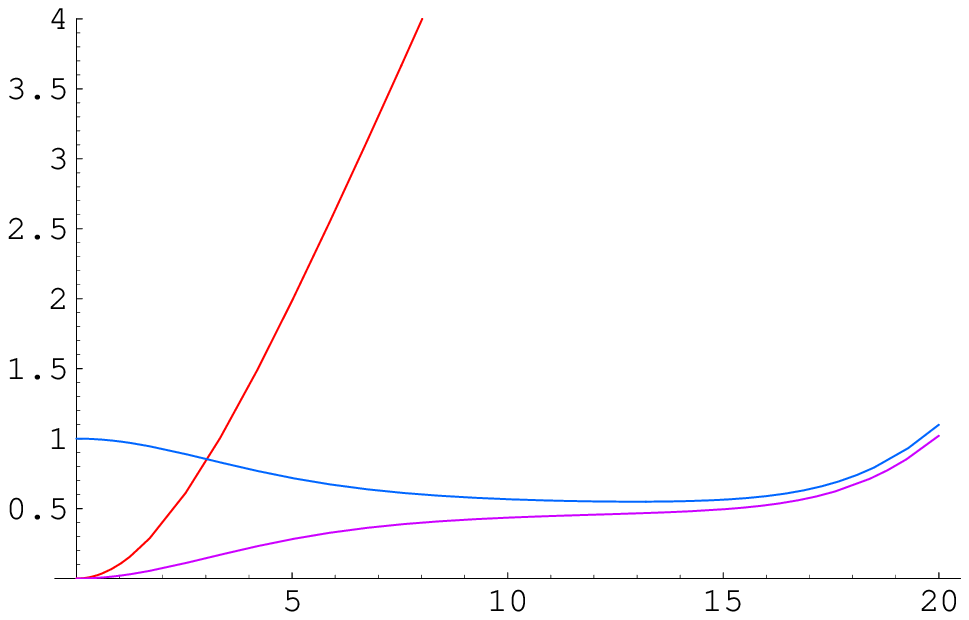} \\
		\includegraphics[width=0.45\textwidth]{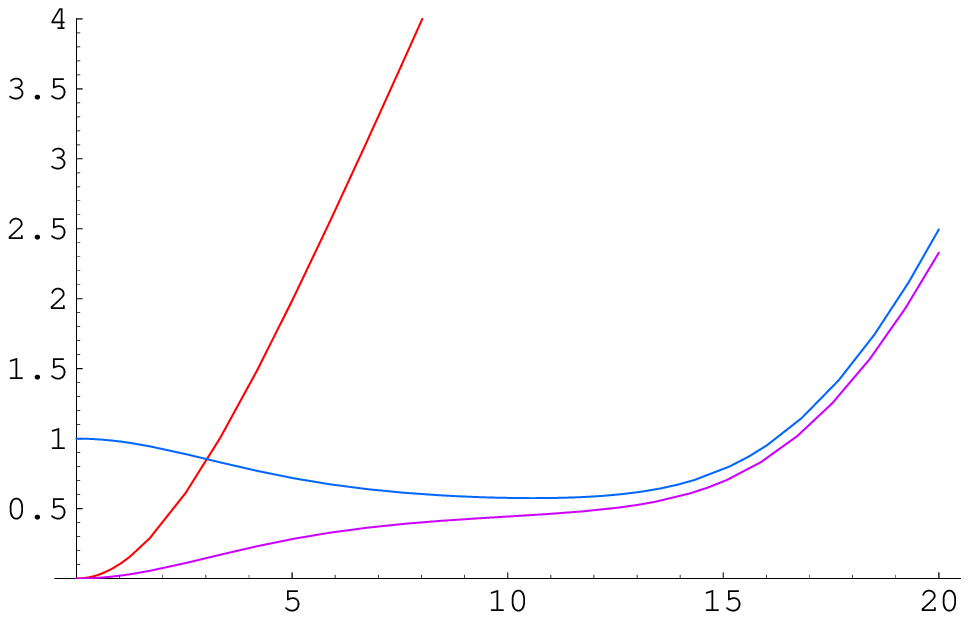} & \includegraphics[width=0.45\textwidth]{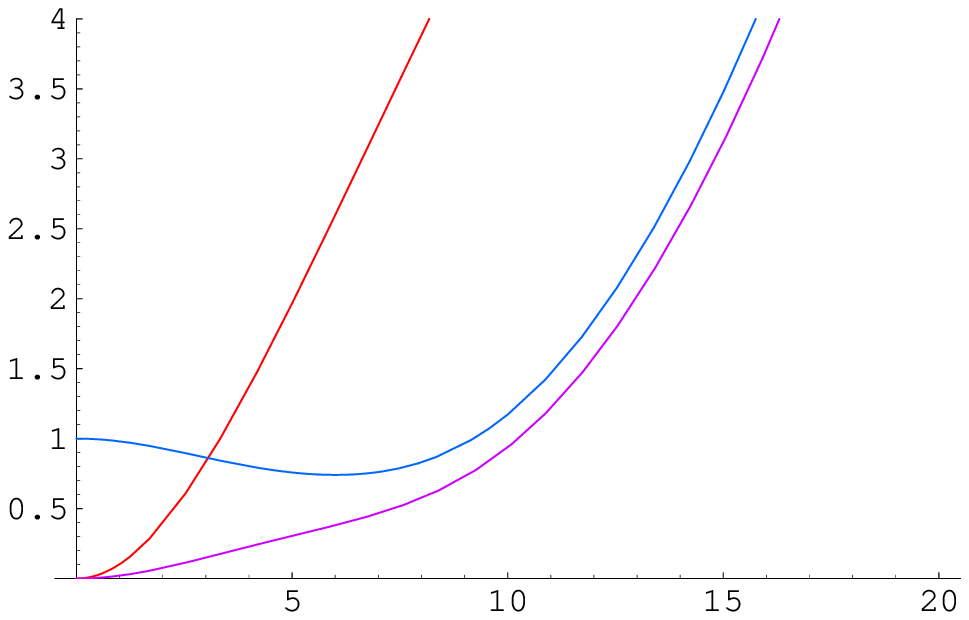}
	\end{tabular}
\caption{Plots of the functions $c^{-1} f_i$  for the $X_{31}$ solution for different values of $c$. $f_1$ is in red, $f_2$ in purple and $f_3$ in blue. The factor $c^{-1}$ is there for normalisation purposes.
From the top left to the bottom right, the values of $c$ are increasing and are $0.125$ (the exact minimum value), 0.125001, 0.12501 and 0.126. The first plot is the Maldacena-Nastase solution.}
\label{fig:Lowc}
\end{figure}

In Figure \ref{fig:Largec} we show plots of the functions for large values of $c$. We see that in the IR the behaviour is that of the space $X_3$. However,  despite starting below $f_3$ (at zero),
 the function $f_1$ 
eventually crosses $f_3$, in agreement with the UV expansions. The crossing point moves
 further and further along  the radial direction as $c$ is increased.
In Figure \ref{fig:Lowc} we plot the functions for values of $c$ close to the minimum.
We see that in the IR the functions are all very close to the special case $c= 1/8$. 
However, when  $c$ is not exactly equal to its minimum value, the functions start to deviate at some point. 
For values of $c$ closer and closer to the special one, there is a larger and larger region where the functions are well approximated by the profiles of the Maldacena-Nastase solution.
\begin{figure}[!ht]
	\begin{tabular}{c c}
		\includegraphics[width=0.45\textwidth]{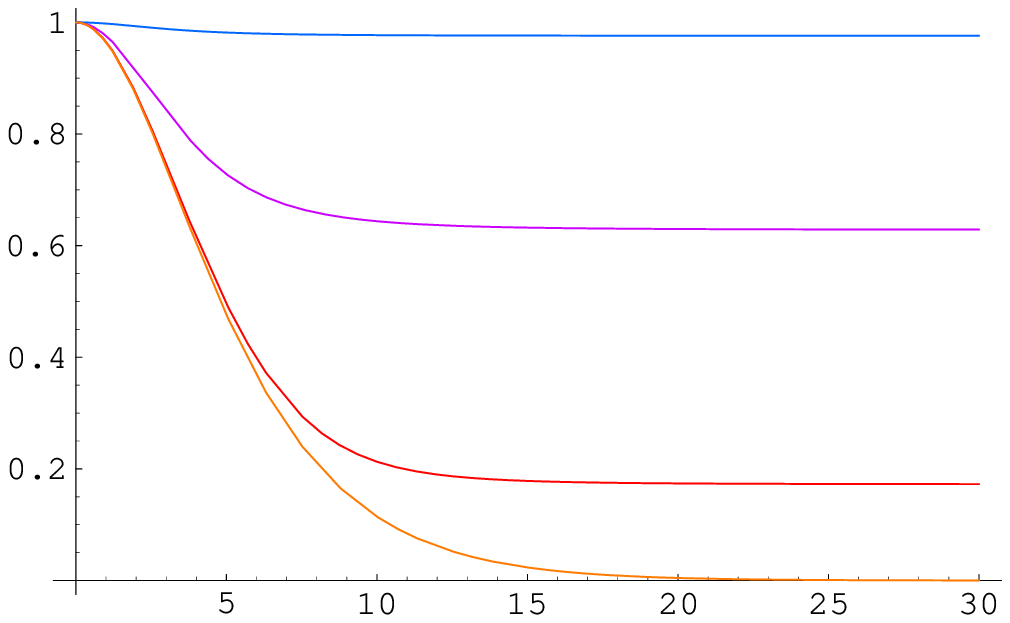} & \includegraphics[width=0.45\textwidth]{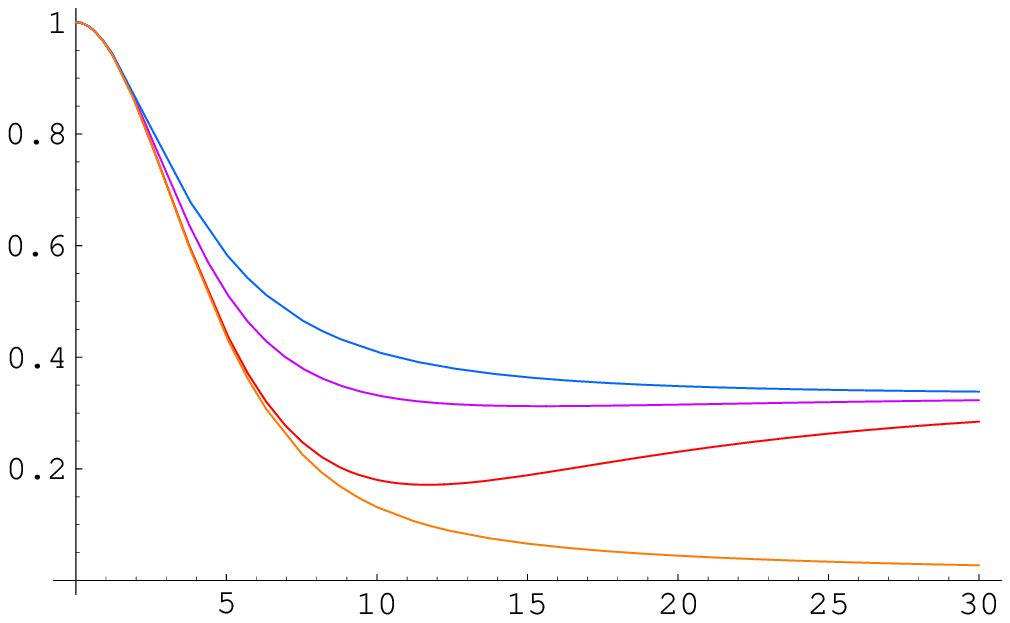}
	\end{tabular}
\caption{On the left are plots of the function $e^{\Phi- \Phi_0}$ for different values of $c$. On the right, plots of the function $\gamma$. The orange plots correspond to the minimum value $c=0.125$, the red ones to $c=0.126$, the purple ones to $c=0.15$ and the blue ones to $c=0.5$. In the Maldacena-Nastase solution at infinity there is a linear dilaton and the $H_3$ vanishes.}
\label{fig:Dilatons}
\end{figure}
Finally, in Figure \ref{fig:Dilatons}, we  plotted the dilaton and the function $\gamma$ for various values of the constant $c$.
We see that the generically $e^{\Phi}$ goes to a constant $e^{\Phi_\infty}$ at infinity, while in the
particular case of Maldacena-Nastase  $e^{\Phi} $ vanishes in the UV.

\subsection{Limits}
\label{limsec}
In this subsection  we analyse two special limits of the one-parameter solutions, namely  $c\to \infty$ and $c\sim 1/8$, respectively.

\subsubsection*{The solution for $c\to \infty:$ $G_2$ holonomy with flux}
\label{sec:G2limit}

 The numerical solutions show that by  increasing
the value of $c$  the solution $X_{31}$ 
looks more and more like the $G_2$ manifold $X_3$.  To see this more precisely,
we will consider an expansion of the functions $f_i$ and $\gamma$ in inverse 
powers\footnote{From \eqref{aposto} it can be checked \emph{a posteriori} that the first few terms reproduce the UV expansions \eqref{genuv}. Therefore the series    
\eqref{eq:cExpansion} is certainly not valid for $c=1/8$. In any case, we only need this for large $c$ here.}
 of $c$ of the form: 
\be \label{eq:cExpansion}
	\begin{aligned}
		f_i &= c \sum_{n=0}^{\infty} \frac{1}{c^n} f_{i(n)} ~,\\
		\gamma &= \sum_{n=0}^{\infty} \frac{1}{c^n} \gamma_{(n)}~.
	\end{aligned}
\ee
For small $t$ it can be checked that this agrees  with the IR expansions \eqref{irexp}.
Then one can solve the system  (\ref{eq:BPSwithf})   order by order in powers of $c^{-1}$. 
There are of course different solutions depending on the boundary conditions and here we will concentrate on the boundary conditions 
already treated in Section \ref{sec:Expansions}. 
For our purpose, we only need the first few orders of the expansion \eqref{eq:cExpansion}.
These read 
\be
	\begin{aligned}
		f_{1(0)} &= f_{2(0)}= \frac{r^3 - r_0^3}{36r}~, \qquad  f_{3(0)} = \frac{2r^3 + r_0^3}{72r} ~,\\
		f_{1(1)} &=   - f_{2(1)}= -\frac{r_0^3 + r_0^2 r + r_0 r^2 - 3 r^3}{12r(r_0^2 + r_0 r + r^2)} ~, \qquad f_{3(1)} = 0 ~,\\
		\gamma_{(0)} &= \frac{r^3 + r_0 r^2 + r_0^2 r + 6 r_0^3}{3r(r^2 + r_0 r + r_0^2)}~,
	\end{aligned}
	\label{aposto}
\ee
where $r$ is a function of the radial coordinate $t$ defined as in Section  \ref{g2metrics:section}, namely 
\be
	\frac{dr}{dt} = \sqrt{1-\frac{r_0^3}{r^3}}~.
\ee
The functions $f_{i(0)}$  at the lowest order in the expansion 
solve a simplified version of  (\ref{eq:BPSwithf})
where  $\gamma_i = 0,	f_1 = f_2$, which are simply the differential equations 
for the $G_2$ holonomy metric $X_3$.  The metric in terms of the expansion \eqref{eq:cExpansion}  reads
\be
	ds^2_7 = M c \left[ ds^2_7 (X_3) + O \left( c^{-1} \right)\right] 
	\label{g2limitmetric}
\ee
thus, at leading order in $c$, the solution looks like the $G_2$ manifold $X_3$ with a very large $S^3$, and $M$ units of $H_3$ flux through it.
One could of course  take $c \rightarrow \infty$ while keeping $Mc$ fixed, by taking $M \rightarrow 0$ at the same time. This is an exact solution, 
where the flux $H_3$ vanishes.

\subsubsection*{The solution for $c\sim 1/8$: $G_2$ holonomy with branes}
\label{branelimit}

Here we show that when $c$ is very close to the minimum value $c=1/8$,  there is a region where the solution $X_{ij}$ looks like  a 
$G_2$ holonomy manifold $X_j$ with $M$ fivebranes wrapped on the non-trivial three-sphere.  The calculation 
is analogous to that appearing in Section A.2 of  \cite{Maldacena:2009mw}.

The solution stays close to the Maldacena-Nastase solution up to large values of $t$. 
To analyse the behaviour  of the solution where it starts departing from this, we consider the following ansatz for an approximate solution 
\be
\begin{aligned}
f_1 &=  \frac{\sqrt{2}}{16} t + \mu_1~,  &~~~~~~~~~~~  f_2 &= \frac{1}{16}  +\mu_2~, \\
f_3 &= \frac{1}{16}  + \mu_3 ~, &~~~~~~~~~~~ \gamma & =   \frac{\sqrt{2}}{2 t} ~.
\end{aligned}
\label{newansa}
\ee
The leading terms are those of the Maldacena-Nastase solution and we 
 require that $\mu_1 \ll  t$,  $\mu_2, \mu_3 \ll 1$. 
Anticipating the form of the metric that we are after, we change coordinates as follows
\be
ds^2_7  =   M \left[ f_2 \left(\frac{1}{8} dy^2 + da_2^2\right) + f_3 da_3^2 + f_1 da_1^2\right]~.
\ee
We could also have taken $f_3$ in front, but since we will find that  $\mu_2=\mu_3$,  this does not matter.  
Then we plug the ansatz \eqref{newansa} into the BPS equations \eqref{eq:BPSwithf} and expand to first order in the $\mu_i$.
The equation for $\gamma'$ is satisfied automatically at leading order. 
Working at large $y$, we can solve the equations for $\mu_i$ and we find
\be
\begin{aligned}
	\mu_1 &= \frac{\beta_1}{2} (e^{\sqrt{2}y/8} -1) + \beta_2 ~,\\
	\mu_2 = \mu_3 &= \beta_1  e^{\sqrt{2}y/8} ~,
\end{aligned}
\ee
where $\beta_1, \beta_2$ are two integration constants. 
We can determine the dilaton with the same precision by considering the ansatz
\be
	\Phi = - \frac{\sqrt{2}}{4}t + \mu_4
\ee
where $\mu_4 \ll t$. Then we find 
\be
	\mu_4 = 8 \beta_1 (e^{\sqrt{2}y/8} - 1) + \Phi_4
\ee
where $\Phi_4$ is another integration constant. Inserting these back into the metric and changing coordinates  as
$r = 4 \sqrt{M \beta_1} e^{\sqrt{2}y/16}$ we find 
\be
\begin{aligned}
	ds^2_7 &\approx \left( 1+ \frac{M}{r^2} \right) \left[ dr^2 + \frac{r^2}{8}  \sum_{i=1}^3\left(\Sigma_i^2  + (\Sigma_i - \sigma_i)^2\right) \right]  
	+ M  \frac{\sqrt{2}}{32}  y \sum_{i=1}^3 \sigma_i^2 ~,\\[2mm]
e^{2(\Phi - \Phi_4) } & \approx 16 \beta_1 e^{-16\beta_1} \left( 1 + \frac{M}{r^2} \right)~.
\end{aligned}
\ee
This is the approximate solution for $M$ fivebranes in flat space, wrapped on the three-sphere parameterised by $\sigma_i$. 
More precisely, we see that the topology is that of $S^3\times \R^4$, where  the three-sphere $C_1$ transverse to the branes
(defined by $\sigma_i=0$) vanishes smoothly. Hence this is the same topology of the  $G_2$ holonomy manifold $X_1$.  The fivebranes then can be wrapping $C_2$ or $-C_3$.

This approximation however requires that  $y$ is large, but at the same time 
$y\ll y_5$, where $y_5$ is  defined by  $\beta_1 = e^{-\sqrt{2}y_5/8}$.
Presumably around $y\sim y_5$ the solution  looks more accurately like $X_1$ 
\cite{Maldacena:2009mw}, but this seems difficult to analyse in the linearised approximation.  
We can also estimate the relation between  $c$ and $y_5$ by extrapolating to zero the value of $f_2+f_3$. This gives
\be
c- 1/8 \approx e^{-\tfrac{\sqrt{2}}{8}y_5}~.
\ee

\subsection{Summary}

In this section we have discussed a set of  gravity solutions 
characterised by a non-trivial parameter $c$. 
The additional parameters of the solutions are the $M$ integral units
of NS three-form flux $H_3$ and the asymptotic value of 
the dilaton $\Phi_\infty$. The constant $\Phi_0$
 is a function of $\Phi_\infty$ and $c$, that may be determined numerically. 
There are six different solutions, exchanged by the action of the 
triality group $\Sigma_3$. In each case the internal  seven-dimensional manifold 
is an  asymptotically conical space,
 with topology $S^3\times \R^4$, that we have denoted  $X_{ij}$, with $i,j=1,2,3$. The base of the asymptotic cone
is the nearly K\"ahler manifold $S^3 \times S^3$ with
 metric $ds^2(Y)= \tfrac{1}{36} (da_1^2 + da_2^2 + da_3^2)$ \cite{Atiyah:2001qf}.
More precisely, the topology of the space $X_{ij}$ is that of the $G_2$ holonomy manifold $X_i$, that we reviewed in Section \ref{g2review}.

In each case the parameter $c$ gives  the size of the non-trivial $S^3$ at the origin, hence
this is analogous to the \emph{deformation}  in the deformed conifold.  
On the other hand, one can also define a resolution  parameter 
by looking at how the metric breaks a $\Z_2 \subset \Sigma_3$ symmetry at large distances. In particular, we have argued that  the 
parameter $1/c$ gives a measure of how much the space $X_{ij}$ deviates from the $X_i$ geometry. Hence, from this point of view,
$1/c$ can be interpreted  as an effective \emph{resolution} parameter. 
In the case of $G_2$ holonomy the moduli space of metrics on $S^3\times \R^4$ has three different branches, meeting at the origin. 
With an abuse of language\footnote{In particular, we do not use these words here in the sense of 
complex or symplectic geometry.}, we can say that the singular $G_2$ cone over $S^3 \times S^3$ may be deformed, resolved or flopped-resolved, with the three
possibilities mutually exclusive. The six solutions that we discussed  may be said to be  deformed \emph{and} resolved, analogously to the resolved deformed conifold geometry  \cite{Butti:2004pk,Maldacena:2009mw}.

When $c$ is very large the solution approaches a $G_2$ holonomy manifold with flux on a large three-sphere.  
When $c$ hits the lower bound $c=1/8$, the $X_{ij}$ geometry becomes a solution of the type  discussed by \cite{Maldacena:2001pb},  
corresponding to the near-brane limit of a large number 
of fivebranes wrapped on the $S^3$ inside a $G_2$ manifold with topology  $S^3\times \R^4$.
These have a finite size three-sphere at the origin, 
 but are asymptotically linear dilaton backgrounds.
When $c$ is very close to the critical value $c=1/8$ the solution stays close to 
 the near-brane Maldacena-Nastase one up to large values of $t$ 
 (see the plots in Figure \ref{fig:Lowc}) and when it starts deviating from this 
 behaviour the geometry  becomes approximately that of the $G_2$ manifold $X_j$ 
 with $M$ fivebranes wrapping the non-trivial three-sphere inside this.

\begin{figure}[ht!]\begin{center}
	\begin{tabular}{c c}
		\includegraphics[width=0.3\textwidth]{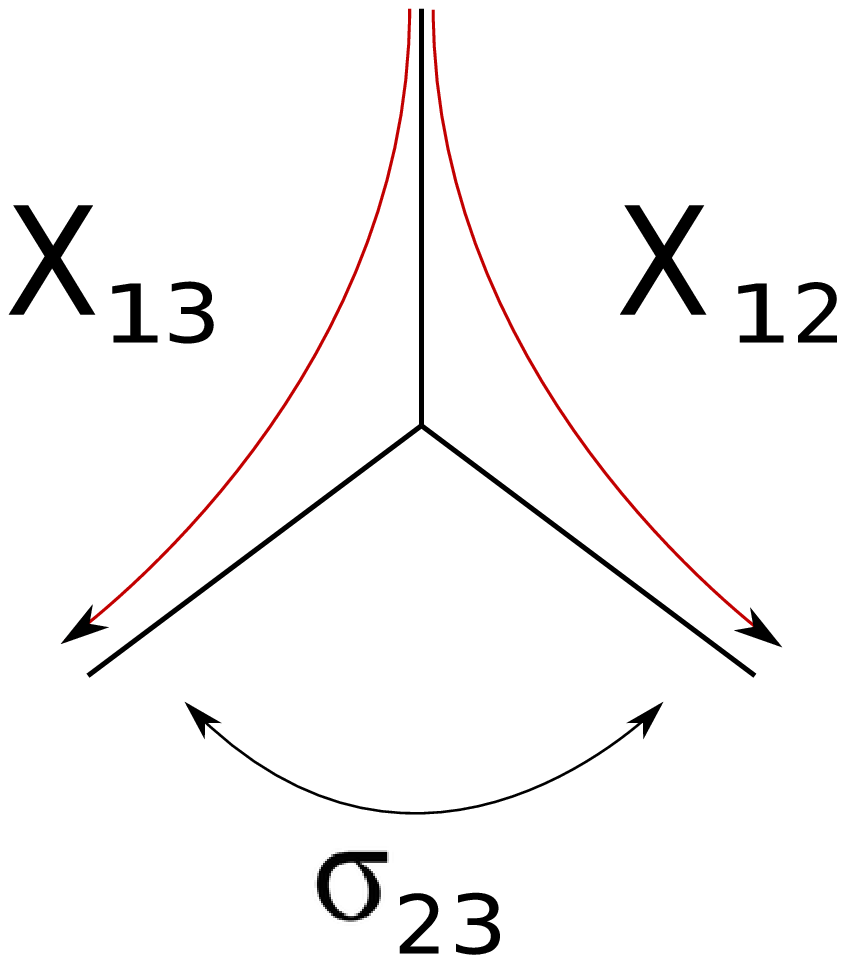} &~~~~~~~~~~~~~ \includegraphics[width=0.3\textwidth]{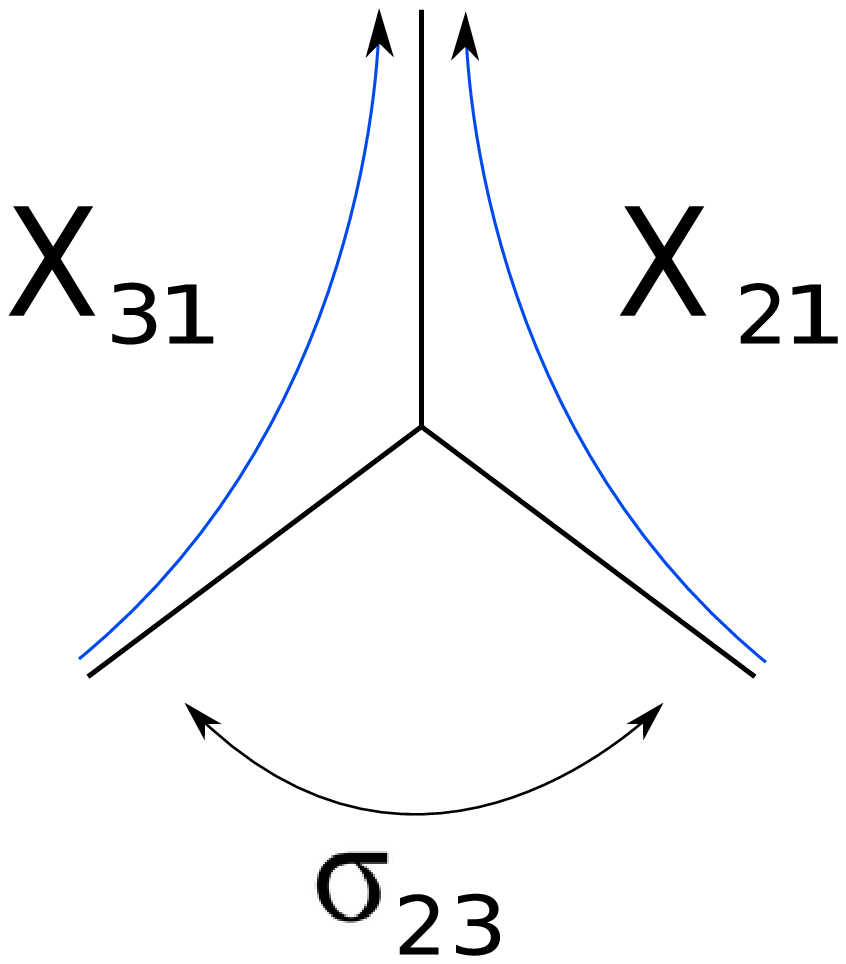} 
	\end{tabular}
\caption{On the left: the solutions $X_{13}$ and $X_{12}$ interpolate continuously 
 between the $G_2$ manifold $X_1$ with flux and the $G_2$ manifolds $X_3$ and $X_2$ with branes, respectively.   The two solutions are related by the $\Z_2$ symmetry $\sigma_{23}$ and both have topology of $X_1\cong S^3\times \R^4$. On the right:  the solution $X_{31}$ interpolates
 continuously  between the $G_2$ manifold $X_3$ with flux and the $G_2$ manifold $X_1$ with branes, while 
the solution $X_{21}$ interpolates between the $G_2$ manifold $X_2$ with flux and the $G_2$ manifold $X_1$ with branes.  
The two solutions are related by the same $\Z_2$ symmetry $\sigma_{23}$, however $X_{31}$  has the topology of $X_3\cong S^3\times \R^4$ while 
$X_{21}$  has the topology of $X_2\cong S^3\times \R^4$.}
\label{g2transfig}
\end{center}
\end{figure}

For each $X_{ij}$ solution the parameter $c$ interpolates between the $G_2$
 holonomy manifold $X_i$ with $M$ units of flux on a large three-sphere, 
and  the $G_2$ holonomy manifold $X_j$ with $M$ fivebranes wrapped on a (different) three-sphere. 
Hence, this may be interpreted as a realisation of a $G_2$
geometric transition, purely in the context of supergravity. Notice that this is different from
 the closely related setup in \cite{Atiyah:2000zz,Atiyah:2001qf}, where
the relevant geometric transition involved D6 branes wrapped in  the conifold, although 
this was embedded in the $G_2$ holonomy context by uplifting to M-theory.

From the point of view of the $G_2$ holonomy manifold $X_1$ (say) with $M$ units of flux through  the three-sphere $C_2  \cong - C_3$,    
the two solutions $X_{12}$ and $X_{13}$ break the $\Z_2$ symmetry under $\sigma_{23}$
in two opposite directions. These look like a ``resolution'' of the 
manifold $X_1$ and its flopped version. The breaking of this $\Z_2$ symmetry is analogous to the breaking of the $\Z_2$ symmetry in the resolved deformed conifold.  On the other hand, 
from the point of view of the branes wrapped on   the three-sphere $C_2  \cong - C_3$ in $X_1$, 
the two solutions $X_{21}$ and $X_{31}$ break the $\Z_2$ symmetry 
generated by $\sigma_{23}$ by ``deforming'' the original  $X_1$ manifold in two different ways.  
In other words,  the $G_2$ geometric transition may proceed from branes in $X_1$ to flux in $X_2$
\emph{or} from branes in $X_1$ to flux in $X_3$.  There is no analogue of this in the conifold case.

Moreover,  depending on which three-sphere of $X_1$ the branes wrap,
 in the geometry after the backreaction  this three-sphere may become contractible or not. 
For example, if the fivebranes were wrapped on $C_2\subset X_1$ and after the geometric transition we have the manifold 
$X_2$ (with flux), the sphere wrapped by the branes is contractible. 
Whereas
if   the fivebranes were wrapped on $C_2\subset X_1$ and after the geometric transition we have the manifold 
$X_3$ (with flux), the sphere wrapped by the branes is \emph{not} contractible.  
This phenomenon has no analogue in the conifold case, where the two-sphere wrapped by the branes always becomes contractible in the backreacted geometry. 
In the context of the discussion in 
\cite{Maldacena:2001pb} these two possibilities led to two different values of the quantity denoted $k_6$, defined as the flux of $H_3$ through the three-sphere wrapped by the branes. 
In particular, for  the basic solution of \cite{Maldacena:2001pb} (with $X_3$ topology) 
it was assumed that the three-sphere wrapped by the branes is $C_2\subset X_1$
and therefore  $k_6=q_2=M$. The second solution 
of \cite{Maldacena:2001pb} (with $X_2$ topology) 
was interpreted as arising from fivebranes still wrapped on 
  $C_2\subset X_1$
  and therefore here  $k_6=q_2=0$. \emph{Cf}. Table \ref{kvalues}.
  This ambiguity may also be understood as related to different gauge choices for the 
   connection on the normal bundle to the wrapped three-sphere. In \cite{Maldacena:2001pb} it was explained how this  corresponds to changing the number of fermionic zero modes on the brane worldvolume with all choices leading to equivalent results for the physical Chern-Simons level of the dual gauge theory, namely $|k|=M/2$.

\section{Type IIA solutions with interpolating $G_2$ structures}
\label{sec:rotation}

In this section we will discuss solutions of Type IIA supergravity 
 of the type $\R^{1,2}\times_w M_7$, where the internal seven-dimensional 
manifold $M_7$ has a $G_2$ structure and there are  various fluxes.
We will then show that starting from a  torsional $G_2$ geometry one can  obtain 
a more general solution,  interpolating between the original solution and a warped $G_2$ holonomy solution. 
The method that we use  is quite general and can be applied to supergravity solutions different from the ones of the previous section.

\subsection{Supersymmetry conditions}

We write the metric ansatz in string frame as
\be
	ds^2_{str} = e^{2\Delta + 2 \Phi/3} \big( dx_{1+2}^2 + ds_7^2 \big)~.
\ee
The solutions  are characterised by a $G_2$ structure on the internal space, 
namely an associative three-form  $\phi$ (and its Hodge dual) and a non-trivial phase $\zeta$. 
The non-zero fluxes are  the  RR four-form $F_{4}$ and the NS three-form $H_3$.  The  equations characterising the geometry 
may be written in the form of generalised calibration conditions \cite{Gutowski:1999tu}, 
and can be obtained straightforwardly  by reducing the equations
presented in  \cite{Martelli:2003ki} from eleven to ten dimensions. Some details of this reduction are presented in Appendix 
\ref{paperzeroeq}. The equations read
\be \label{eq:GeneralBPS}
	\begin{aligned}
		d \big( e^{6\Delta} *_7 \phi \big) &= 0~, \\
		\phi \wedge d \phi &= 0 ~,\\
		2 d \zeta - e^{-3\Delta} \cos \zeta d \big( e^{3\Delta} \sin \zeta \big) &= 0~,\\
                 d \big( e^{2\Delta+ 2\Phi/3} \cos \zeta \big) &= 0 ~.\\
	\end{aligned}
\ee
In addition, the fluxes are determined as follows
\be
	\begin{aligned}
		H_3 &= \frac{1}{\cos^2 \zeta} e^{-4\Delta + 2\Phi/3} *_7 d \big( e^{6\Delta} \cos \zeta \phi \big)~, \\
		F_{4} &= \vol_3 \wedge d \big( e^{3\Delta} \sin \zeta \big) + F_4^{\mathrm{int}}~,\qquad F_4^{\mathrm{int}} = - \frac{\sin \zeta}{\cos^2 \zeta} e^{-3\Delta} d \big(e^{6\Delta} \cos \zeta \phi \big)~.
	\end{aligned}
\label{generalfluxes}
\ee
Notice the  relation
\bea
\sin \zeta e^{\Delta-2\Phi/3} H_3 + *_7 F_4^{\mathrm{int}}  =  0 ~.
\label{selfish}
\eea
 From the results of  \cite{Martelli:2003ki} we have that any solution to these conditions,
supplemented by the Bianchi identities $dH_3 = dF_4 = 0$, solves also the equations of motion. 
This geometry is the  $G_2$ analogue of the interpolating $SU(3)$ structure geometry discussed in  \cite{Minasian:2009rn,Gaillard:2010qg}.
Notice that this case is not contained in the equations presented in 
\cite{Haack:2009jg}, which instead describe an $SU(3)$ structure in seven dimensions.
Although  the 
ansatz for the bosonic fields in the latter reference is equivalent to ours, the ansatz for the Killing spinors does not allow the structure that we
are discussing here. The interested reader can find a discussion of spinor ansatze in Appendix  \ref{paperzeroeq}.

The conditions \eqref{eq:GeneralBPS},  \eqref{generalfluxes} include the Type I torsional geometries 
 as a special case, which are obtained simply  setting $\zeta=\pi$. The warp factor is related\footnote{An integration constant can always be reabsorbed in a scaling of the coordinates.}  to the dilaton as $e^{2\Phi} = e^{-6\Delta}$ so that the ten-dimensional metric in string frame is unwarped. 
The limit $\cos \zeta \to 0$ is slightly singular, since in this case the $G_2$ structure in eleven dimensions from which our equations have been obtained
breaks down\footnote{In particular, the one-form $K$  does not exist. See Appendix \ref{paperzeroeq} and \cite{Martelli:2003ki}.}. 
However, going back to the equations in \cite{Martelli:2003ki}, one can see
that in this limit the internal eight-dimensional geometry  becomes a warped Spin(7) manifold with self-dual flux 
\cite{Becker:2000jc}. A careful analysis then 
shows that in the  $\cos \zeta \to 0$ limit we obtain the warped $G_2$ holonomy solutions derived in \cite{Cvetic:2000mh}.  The warp factor is again related to the dilaton
\bea
e^{4\Phi} = e^{-3\Delta} \equiv h 
\eea
and rescaling the internal metric as $ds^2_7 = h d \hat s^2_7$ the 
full metric becomes 
\be
	ds^2_{str} = h^{-1/2}dx_{1+2}^2 + h^{1/2}d\hat s_7^2 ~
\ee
where the rescaled metric now has $G_2$ \emph{holonomy}, namely $d\hat \phi = d \hat *_7 \hat \phi = 0$. 
Taking directly the limit on the relation 
\eqref{selfish} gives  $ H_3 + \hat *_7 F_4^{\mathrm{int}}  =  0$. Hence the four-form flux can be written  as 
\be
	\begin{aligned}
			F_{4} &= \vol_3 \wedge d  h^{-1} - \hat *_7 H_3 ~.
	\end{aligned}
\ee
The equation of motion for $F_4$ implies that the warp factor is harmonic with respect to the  $G_2$ metric, namely
\be
\mathrm{\raisebox{1.1mm}{$\hat{} $}} \!\! \Box_7 h = - \frac{1}{6} H_3^2~.
\ee

\subsection{Solution generating method}
\label{sec:method}

We now discuss two different methods to generate solutions of the equations presented above, starting from a solution of the Type I torsional system. 
One method, analogous to  the procedure discussed in \cite{Maldacena:2009mw}, involves a 
simple chain of  dualities. Another method  exploits the form of the supersymmetry conditions. 
We will refer to this second method as ``rotation'' \cite{Minasian:2009rn,Gaillard:2010qg}.

\paragraph{Dualities}

We start with a solution of \eqref{torsional}. The non-trivial fields are the  dilaton $\Phi$, 
the three-form flux $H_3$ and the metric  $ds^2_{str} = dx_{1+2}^2 + ds_7^2$.
First we uplift to eleven dimensions. We rescale the new eleventh dimension by a constant factor $e^{-\Phi_{\infty}}$, 
boost along $x_{11}$ with parameter $\beta$, and finally undo the rescaling of $x_{11}$. This gives the transformation 
\bea
t ~\to ~ \cosh\beta t - \sinh \beta e^{\Phi_{\infty}} x_{11}~, ~~~~~~~~~~~ x_{11} ~\to ~ - \sinh\beta e^{-\Phi_{\infty}} t + \cosh \beta x_{11}~.
\eea
Then we reduce back to Type IIA along the transformed $x_{11}$ and we perform two T-dualities along the two spatial directions of the $\R^{1,2}$ part.
At the level of brane charges, the steps in the  transformation may be summarised as
\begin{center}
NS5 ~~$\to$~~M5   ~~$\to$~~ M5, $p_{KK}$ ~~$\to$~~ NS5, D0 ~~$\to$~~ NS5, D2
\end{center}
Notice that a non-zero magnetic $\hat C_3$ field will be generated in the process. The dualities above result in the following Type IIA solution
\be
\begin{aligned}
d\hat{s}^2_{str} ~ &= ~ h^{-1/2} dx^2_{1+2} + h^{1/2} ds^2_7 ~,~~~~~~~~~~~~~~~  h = 1+ \sinh^2 \beta  (1 - e^{-2(\Phi- \Phi_\infty)}) ~,\\
 \hat H_3 ~ & = ~ \cosh \beta H_3~, ~~~~~~~~~~~~~~~~~~~~~~~~~~~e^{2\hat\Phi} ~ = ~  e^{2\Phi} h^{1/2}~,\\
\hat F_4 ~ &= ~ - \frac{e^{-\Phi_{\infty}}}{\tanh \beta}  \vol_3\wedge d (h^{-1}) + \sinh \beta e^{\Phi_{\infty}} e^{-2\Phi}*_7 H_3  ~, 
\end{aligned}
\label{dualsolu}
\ee
where here the hatted quantities denote the new solution while the unhatted ones denote the initial solution. 
Notice that in contrast to the case in \cite{Maldacena:2009mw} the dilaton is changed in the transformation. This can be understood because
here the procedure introduces D2 branes, to which the dilaton couples.
Notice also that we need $h>0$,  which imposes  $e^{2\Phi-2\Phi_{\infty}} > \tanh^2 \beta$. Thus the transformation may be applied only if in the initial solution the dilaton is a bounded function. We can write the  transformed fluxes as
\be
\begin{aligned}
\hat H_3 &=  - \cosh \beta e^{2\Phi} *_7 d(e^{-2\Phi} \phi) \\[2mm]
\hat F_4 &=  - \frac{e^{-\Phi_{\infty}}}{\tanh \beta}  \vol_3\wedge d h^{-1} -  \sinh \beta e^{\Phi_{\infty}} d(e^{-2\Phi}\phi ) 
\end{aligned}
 \label{eq:RotatedSolution}
\ee
from which we can read off the internal $\hat C_3$ field in terms of the associative three-form $\phi$, namely
\bea
\hat C_3 = - \sinh \beta e^{\Phi_{\infty}-2\Phi} \phi~.
\eea 
From these expressions it is clear that the Bianchi identities of the initial solution imply the ones of the transformed solution. 
In principle this method may be  applied also to non-supersymmetric solutions. 

\paragraph{Rotation}

The same transformation can be done directly on the supersymmetry equations, without doing any dualities. 
One advantage of this method is for example
that it is applicable to configurations with sources \cite{Gaillard:2010qg}. 
 Suppose that $\Phi^{(0)}=-3\Delta^{(0)}$ and a three-form $\phi^{(0)}$ are a solution of the system \eqref{torsional}. 
Then one can define 
\be
	\begin{aligned}
		\hat \phi &= \left(\frac{\cos \zeta}{c_1} \right)^3 \phi^{(0)} ~,\\
		e^{2\hat \Phi} &= \frac{\cos \zeta}{c_1} e^{2\Phi^{(0)}} ~,\\
		e^{3 \hat \Delta} &= \left(\frac{c_1}{\cos \zeta} \right)^2 e^{-\Phi^{(0)}}~,
	\end{aligned}
\ee
and a new seven-dimensional metric $d\hat s^2_7= c_1^{-2} \cos^2 \zeta ds_7^{(0)2}$. 
It is easy to check that the new  quantities $\hat \Phi$, $\hat \Delta$ and $\hat \phi$ are a solution of the first three equations of the general system \eqref{eq:GeneralBPS}. The fourth one can be solved and it gives a relation between $\zeta$ and the dilaton of the original solution:
\be
	\sin \zeta =  c_2 e^{-\Phi^{(0)}} ~.
\ee
Here $c_1$ and $c_2$ are integration constants.
The rotated background in terms of unrotated quantities reads
\be
	\begin{aligned}
		d\hat s^2_{str} &= h^{-1/2} dx_{1+2}^2 + h^{1/2} ds_7^{(0)2}~,  \qquad \quad ~~~~h = \frac{1}{c_1^2} \left( 1 -  c_2^2  e^{-2\Phi^{(0)}} \right)~,\\
		\hat H_3 &= \frac{1}{c_1} e^{2\Phi^{(0)}} *_7^{(0)} d \big( e^{-2\Phi^{(0)}} \phi^{(0)} \big) ~, ~~~~~~~~~~~~~~   e^{2\hat \Phi} =  e^{2\Phi^{(0)}}  h^{1/2}~,\     \\
		\hat F_{4} &= \frac{1}{c_2} \vol_3 \wedge d h^{-1} - \frac{c_2}{c_1} d \big(e^{-2\Phi^{(0)}} \phi^{(0)} \big)~.
	\end{aligned}
\label{rotatedsolu}
\ee  
In order to match the result of this method to the previous one, one has to identify 
\be
	c_1 ~ = ~  -\frac{1}{\cosh \beta}~, ~~~~~~~~~~~~~~ c_2 ~=~ -e^{\Phi_\infty}\tanh \beta~.
\ee
As before, the Bianchi identities of the general solution  follow immediately from the ones of the unrotated solution.

\subsection{Deformations of the warped $G_2$ holonomy solutions}

We can now apply the transformation above to the solutions 
of Section  \ref{solutions:sec}. Notice that indeed  in those solutions 
the dilaton was bounded from below.
For each solution of Section \ref{solutions:sec} we then obtain a one-parameter
family of solutions of Type IIA supergravity, with D2 brane charge and an internal $C_3$ field.  
The background is simply obtained by plugging the solutions
of Section \ref{solutions:sec} into the equations \eqref{dualsolu} or \eqref{rotatedsolu}. 
Notice that the warp factor $h$ in \eqref{dualsolu} goes to one at infinity. 
However, for AdS/CFT applications, one would like to take 
a decoupling limit in which the warp factor goes to zero 
at infinity. In this way the asymptotically Minkowski region is removed and replaced by a boundary. We will be more precise about the asymptotics  momentarily.  To proceed, first recall that one should quantise the transformed three-form $\hat H_3$ as  
\bea 
\tilde M = \frac{1}{4\pi^2} \int_{S^3} \hat H_3 = M \cosh \beta \in \N
\eea
where $S^3$ is the appropriate non-trivial three-sphere in each case. Then  rescaling the Minkowski coordinates as
\bea
x^\mu \to \left(\frac{\tilde M \cosh \beta}{c }\right)^{1/2} x^\mu~,
\eea 
in the limit $\beta \to \infty$, keeping $\tilde M$ fixed,  the metric is finite and reads 
\be
d\hat s^2_{str} = \tilde M \left[ \tilde h^{-1/2} c^{-1} dx_{1+2}^2 + \tilde h^{1/2} d\tilde s_7^{2} \right]~.
\ee
Here $d\tilde s^2_7$ does not have a factor of $M$ 
and the new warp factor $\tilde h = 1- e^{-2(\Phi - \Phi_\infty) }$  goes to zero at infinity.
The factor of $c$ makes sure that the asymptotic form of the metric is independent of $c$
 and in addition   will allow us to take the further limit $c\to \infty$. 
From the expressions in \eqref{dualsolu} we see 
that this limit is problematic for the transformed $\hat F_4$
and dilaton $\hat \Phi$.  
To obtain a finite limit we also send $e^{\Phi_\infty} \to 0$, while keeping fixed 
\be
e^{2\Phi_\infty}\sinh\beta =  c~.
\ee
The factor $c$ on the right-hand side is again inserted to allow to take a further 
$c\to \infty$ limit in the solution. Now taking $\beta \to \infty$ the solution is then completed
with\footnote{In the expressions below $\Phi_\infty$ enters only in the 
combination $\Phi - \Phi_\infty$, which is finite in the limit.} 
\be
	\begin{aligned}
	e^{2\hat \Phi} & =    c \, e^{2 (\Phi -\Phi_\infty)}   \tilde h^{1/2} ~, ~~~~~~~~~\hat H_3 ~= - \tilde M e^{2(\Phi -\Phi_\infty)} \tilde *_7 
d ( e^{-2  (\Phi   -\Phi_\infty   )}\tilde \phi )  ~,\\[2mm]
		\hat F_{4} &=  - \tilde M^{3/2} \left[ 
		c^{-2} \vol_3 \wedge d \tilde h^{-1} + c^{-1/2}
		d \big( e^{-2(\Phi -\Phi_\infty)}  \tilde \phi  \big)   \right]~.
	\end{aligned}
\ee  
Here tildes on $\tilde *_7$ and $\tilde \phi$ indicate that the expressions are computed with the metric $d\tilde s^2_7$.
We can now show that in this solution the limit $c\to \infty$ 
gives a solution of the type found in \cite{Cvetic:2000mh}.
Firstly, as we saw in Section \ref{sec:G2limit}, for large $c$ the metric 
for each  $X_{ij}$ solution reads
\be
d\hat s^2_{str} = \tilde M \left[ \tilde h^{-1/2} c^{-1} dx_{1+2}^2 + \tilde h^{1/2} c
\left( d  s_7^{2} (X_i) + O(c^{-1}) \right) \right]~.
\ee
From the differential equation for the dilaton we find that
\be
	\Phi' = c^{-2} \left( \tfrac{1}{2} H' + O \left( c^{-1} \right) \right)
\ee
where $H$ is the warp factor of the solution found in \cite{Cvetic:2000mh}, which 
reads
\be
	\begin{aligned}
		H &=\, \frac{3 (r_0 + r)}{4 r_0^3 r^3 (r^2 + r_0 r + r_0^2)^3} \big(16 r^7 + 24 r_0 r^6 + 48 r_0^2 r^5 + 47 r_0^3 r^4 + 54 r_0^4 r^3 \\
& + 36 r_0^5 r^2 +18 r_0^6 r + 9 r_0^7 \big) + \frac{8 \sqrt{3} }{r_0^4} \arctan \frac{2r+r_0}{\sqrt{3} r_0} + q~.
	\end{aligned}
\label{popesd}
\ee
 $q$ is an integration constant and taking
$q = -\frac{4 \sqrt{3} \pi}{r_0^4}$ we have that  $H \sim 81/(4r^4)$ when $r \rightarrow \infty$.
Solving for the dilaton in an expansion  in $c^{-1}$ we find 
\be
	e^{2(\Phi -\Phi_\infty)} =  1 + c^{-2} H + O \left(c^{-3} \right)~.
\ee
Notice that although this was obtained in \cite{Cvetic:2000mh} for the $G_2$ holonomy
 metric  on $X_{1}$, it follows from our discussion in Section \ref{limsec}, that this expression is invariant under 
 $\Sigma_3$ and hence the same function $H$ in \eqref{popesd}
appears for any $X_i$.
Thus taking the limit $c\to \infty$ on a  $X_{ij}$ solution,  gives the following solution
 \be
	\begin{aligned}
	d\hat s^2_{str} & = \tilde M \left[ H^{-1/2} dx_{1+2}^2 + H^{1/2} d  s_7^{2} (X_i)  \right]~,~~~~~~~~e^{2\hat \Phi}  =   H^{1/2} ~, \\[2mm]
		\hat F_{4} &=  - \tilde M^{3/2} \left[ 
		\vol_3 \wedge d \tilde H^{-1} -  *_7 L_3  \right] ~,~~~~~~~~~~~~\hat H_3  = L_3~,
	\end{aligned}
\label{betainf}
\ee  
where $L_3$ is a harmonic three-form\footnote{This can be extracted from the $c\to \infty$ limit of 
$c^{-1/2}\tilde *_7d \tilde \phi$.} on $X_i$. This is precisely 
the warped $G_2$ solution presented in  \cite{Cvetic:2000mh}. Notice that asymptotically the string frame metric goes to 
 AdS$_4\times Y$, where $Y\cong S^3\times S^3$, however the dilaton vanishes like $e^{2\hat \Phi}\sim 9/(2r^2)$. In fact, by setting 
$r_0=0$ we have the exact solution with metric
\be
d\hat s^2_{str}  = \frac{9}{2}\tilde M \left[ ds^2( \mathrm{AdS}_4) + ds^{2}(Y)   \right]~,
\label{nksol}
\ee
$e^{2\hat \Phi}= 9/(2r^2)$ and non-trivial $F_4$ and $H_3$ fluxes. However, the solution does not have conformal symmetry
because the dilaton depends on the radial coordinate $r$. 
 In \eqref{nksol} we can replace $Y\cong S^3\times S^3$ with another nearly K\"ahler metric, provided there exists the appropriate harmonic 
three-form  $L_3$ on the $G_2$ cone, thus obtaining a solution generically preserving the same amount of supersymmetry.
These metrics are in fact solutions of massive Type IIA supergravity  \cite{Behrndt:2004km}. 
This is a curious fact that might be relevant for AdS/CFT applications \cite{Gaiotto:2009mv}.

In conclusion, for any solution of the type of   \cite{Cvetic:2000mh}, arising from configurations of 
D2 branes and fractional NS5 branes transverse to a $G_2$ manifold $X_i$,
we have constructed a one-parameter family of deformations, with the same AdS$_4\times Y$ asymptotics. 
These are  analogous to the baryonic branch deformations 
\cite{Butti:2004pk} of the Klebanov-Strassler solution   \cite{Klebanov:2000hb}. 
In particular, they break the $\Z_2\subset \Sigma_3$ symmetry of a $G_2$ holonomy manifold $X_i$.

\section{Discussion}

In this paper we have discussed various supergravity solutions related to configurations of 
fivebranes wrapping a three-sphere in a $G_2$
holonomy manifold $X_i\cong S^3\times \R^4$. Our basic solutions 
are examples of torsional $G_2$ manifolds \cite{Gauntlett:2002sc}
and comprise some cases previously 
studied in \cite{Acharya:2000mu,Maldacena:2001pb,Canoura:2008at}. 
There are six solutions characterised by a non-trivial parameter. As we change this parameter, each solution interpolates between a $G_2$ manifold with (NS5 or D5) branes on a three-sphere and a distinct $G_2$ manifold 
with (NS or RR) flux on a different three-sphere. This is then an explicit realisation of a geometric 
transition between a pair of $G_2$ manifolds, analogous to the version of the conifold transition described in \cite{Maldacena:2009mw}. 
The six solutions pairwise connect the three branches of the classical 
moduli space of $G_2$ holonomy metrics on $S^3\times \R^4$ \cite{Atiyah:2001qf}. It would be
 interesting to see if 
the picture that we discussed, which is purely classical, may be related to a ``large $N$ duality'' similar to 
\cite{Vafa:2000wi}.

From each of the basic solutions we constructed new Type IIA backgrounds with D2 brane charge by employing a simple generating 
method applicable to a class of geometries with interpolating $G_2$ structure.
The solutions constructed in this way are one-parameter deformations of the solutions presented in \cite{Cvetic:2000mh}, 
corresponding to  D2 branes and fractional NS5 branes transverse to the $G_2$ manifold $S^3\times \R^4$. 
Therefore, they are  analogous to the baryonic branch deformation \cite{Butti:2004pk} of the Klebanov-Strassler solution \cite{Klebanov:2000hb}.

Based purely on supergravity considerations,  it is natural to  expect a close relation
between the ${\cal N}=1$ Chern-Simons theory discussed in  
\cite{Maldacena:2001pb} and the ${\cal N}=1$ three-dimensional 
field theory dual to the solutions above. 
Let us conclude with some speculations about the field theory duals.
First of all, the existence of a finite size $S^3$ in the geometry suggests that the IR field theory 
should be confining, as in \cite{Maldacena:2001pb}.
A standard computation of the number of D2 branes $N$ shows that this 
is running,  and vanishes in the IR,  as\footnote{Here the running is not logarithmic, but we have $N(t)\sim M^2t $ at large $t$.} 
 the number of D3 branes in  \cite{Klebanov:2000hb}.
This suggests that the three-dimensional field theory may be a quiver with gauge group 
$U(N)\times U(N+M)$. Moreover, we expect that 
the three-form flux $H_3$ will induce Chern-Simons terms, like in  \cite{Maldacena:2001pb}. We also have a 
running $C_3$ field on a three-sphere at infinity, analogous to the $B$ field in 
 \cite{Klebanov:2000hb}, with $k=\int C_3 \sim M t $, suggesting the relation $N = kM$.  
A possible scenario is therefore that the solution of \cite{Cvetic:2000mh} describes 
a ``cascading'' three-dimensional quiver, which in the ``last step''
becomes the $U(M)_{M/2}$ theory of  \cite{Maldacena:2001pb}.  
However, an important caveat is that our solutions are related to 
 NS5 branes in Type IIA, while the discussion in  \cite{Maldacena:2001pb} applies to configurations of Type IIB NS5 branes. 
Nevertheless, the relation between the various solutions based on the conifold and the solutions we 
discussed here
 indicates 
 that a precise connection between the four-dimensional ``parent'' 
field theories and the three-dimensional field theories should exist, along the lines of \cite{Aganagic:2009zk,Martelli:2009ga}. 
In particular, we can uplift our solutions to M-theory and subsequently reduce along a $U(1)$
 inside the non-trivial geometry, thus obtaining
solutions with the topology of the deformed or resolved conifold \cite{Atiyah:2000zz} (times $S^1$).

Another relationship between the $U(M)_{M/2}$ Chern-Simons theory and the putative field theory dual to  \cite{Cvetic:2000mh} is suggested by the 
one-parameter deformations of \cite{Cvetic:2000mh} that we described. In particular, we saw how in a certain regime of the  parameter ($c\sim 1/8$)
the solution becomes close to the fivebrane geometry. The presence of a large $C_3$ field on the three-sphere wrapped by these branes suggests that 
perhaps the relevant field theory is the theory on fivebranes wrapped on a fuzzy three-sphere \cite{Maldacena:2009mw}.

We leave the investigation of these ideas for future work.

\section*{Acknowledgements}

We are very grateful to Juan Maldacena, Carlos  N\'u\~nez, Johannes Schmude and James Sparks
for discussions and useful comments.  We also thank Diego Rodriguez-Gomez for comments and collaboration on related topics. 
D. M. is partially supported by an EPSRC Advanced Fellowship EP/D07150X/3.

\appendix 

\section{$SU(2)$ invariant one-forms}
\label{appdetails}

Consider three elements $\aone, \atwo, \athree \in SU(2)$ obeying the constraint $ \aone \atwo \athree  = 1$.
We define the following $SU(2)^3$ Lie-algebra valued one-forms
\be
\begin{aligned}
a_1^{-1}da_1 & \equiv  \frac{i}{2} \alpha_i \tau_i\\ 
a_2 da_2^{-1} & \equiv  \frac{i}{2} \beta_i \tau_i\\ 
a_3^{-1}da_3 & \equiv  - \frac{i}{2} \gamma_i \tau_i
\end{aligned}
\ee
where $\tau_i$ are Pauli matrices. We can invert these obtaining
 \be
\begin{aligned}
\alpha_i & = - i\mathrm{Tr}[\tau_i a_1^{-1}da_1]\\ 
\beta_i & = -i \mathrm{Tr}[\tau_i a_2 da_2^{-1}]\\
 \gamma_i & = \,  i \mathrm{Tr}[\tau_i a_3^{-1}da_3]
\end{aligned}
\label{invoneforms}
\ee
Parameterising the group elements explicitly in terms of angular variables as 
\be
	\begin{aligned}
		a_1 &= e^{-i \phi_1 \tau_3 /2} e^{-i \theta_1 \tau_1 /2} e^{-i \psi_1 \tau_3 /2} \\
		a_2 &= e^{i \psi_2 \tau_3 /2} e^{i \theta_2 \tau_1 /2} e^{i \phi_2 \tau_3 /2} \\
		a_3 &= a_{2}^{-1} a_{1}^{-1} = e^{-i \phi_2 \tau_3 /2} e^{-i \theta_2 \tau_1 /2} e^{-i (\psi_2 - \psi_1) \tau_3 /2} e^{i \theta_1 \tau_1 /2} e^{i \phi_1 \tau_3 /2}
	\end{aligned}
\label{defabc}
\ee
after some computation we get
\be
\begin{aligned}
\alpha_1 + i \alpha_2 & = - e^{-i\psi_1} (d\theta_1+ i \sin\theta_1 d\phi_1)~,   ~~~~~~~~~\alpha_3 = - (d\psi_1 + \cos\theta_1 d\phi_1)~,\\
\beta_1 + i \beta_2 & = - e^{-i\psi_2} (d\theta_2+ i \sin\theta_2 d\phi_2) ~,  ~~~~~~~~~\beta_3 = - (d\psi_2 + \cos\theta_2 d\phi_2)~.
\end{aligned}
\ee
Notice $\alpha_i$ and $\beta_i$ are $SU(2)$ left-invariant one-forms,  obeying 
\bea
d\alpha_3 = + \alpha_1 \wedge \alpha_2 ~,~~~~~~~~ d\beta_3 = + \beta_1 \wedge \beta_2 ~,
\eea
and cyclic permutations.  We can also define the following Lie-algebra valued one-forms
  \be
\begin{aligned}
a_1da_1^{-1} & \equiv  \frac{i}{2} \tilde \alpha_i \tau_i\\ 
a_2^{-1} da_2 & \equiv  \frac{i}{2} \tilde \beta_i \tau_i
\end{aligned}
\ee
A similar computation gives
\be
\begin{aligned}
\tilde \alpha_1 + i \tilde \alpha_2 & = e^{i\phi_1} (d\theta_1- i \sin\theta_1 d\psi_1)~,   ~~~~~~~~~\tilde\alpha_3 =  d\phi_1 + \cos\theta_1 d\psi_1~,\\
\tilde \beta_1 + i \tilde \beta_2 & = e^{i\phi_2} (d\theta_2 - i \sin\theta_2 d\psi_2) ~,  ~~~~~~~~~\tilde \beta_3 =  d\phi_2 + \cos\theta_2 d\psi_2~.
\end{aligned}
\ee
These are $SU(2)$ right-invariant one-forms, obeying
\bea
d\tilde\alpha_3 = + \tilde\alpha_1 \wedge \tilde\alpha_2 ~,~~~~~~~~ d\tilde\beta_3 = + \tilde\beta_1 \wedge \tilde\beta_2 ~,
\eea
and cyclic permutations.  Computing the $\gamma_i$ we obtain 
\bea
- \gamma_i = \tilde \alpha_i + M_{ij}\beta_j
\eea
where $M_{ij}$ is the following $SO(3)$ matrix
{\small \be
M_{ij}=\left(
\begin{array}{ccc}
 \text{cos}\phi_1  \text{cos}\psi_1 -\text{cos}\theta_1  \text{sin}\phi_1  \text{sin}\psi_1  & -\text{cos}\theta_1  \text{cos}\psi_1  \text{sin}\phi_1 -\text{cos}\phi_1  \text{sin}\psi_1 & \text{sin}\theta_1  \text{sin}\phi_1  \\
 \text{cos}\psi_1  \text{sin}\phi_1 +\text{cos}\theta_1  \text{cos}\phi_1  \text{sin}\psi_1  & \text{cos}\theta_1 \text{cos}\phi_1  \text{cos}\psi_1 -\text{sin}\phi_1  \text{sin}\psi_1  & -\text{cos}\phi_1  \text{sin}\theta_1  \\
 \text{sin}\theta_1  \text{sin}\psi_1  & \text{cos}\psi_1  \text{sin}\theta_1  & \text{cos}\theta_1 
\end{array}
\right)\\
\ee
}
We note the following identities
\bea
\sum_i \alpha_i^2 = \sum_i \tilde \alpha_i^2~,~~~~~~~~~~ \sum_i \beta_i^2 = \sum_i \tilde \beta_i^2~,
\eea
and 
\bea
\sum_i \gamma_i^2 = \sum_i (\alpha_i -\beta_i)^2 ~.
\eea
To prove the latter we have to use $M_{ij}M_{ik}= \delta_{ik}$ and  $\alpha_i = - M_{ji}\tilde \alpha_j$.
We identify the above with the (left-invariant) one-forms $\sigma_i$ and $\Sigma_i$ used in the main text
\be
\begin{aligned}
\sigma_i & = -\alpha_i = i\mathrm{Tr}[\tau_i a_1^{-1}da_1]\\ 
\Sigma_i & = - \beta_i  = i \mathrm{Tr}[\tau_i a_2 da_2^{-1}]
\end{aligned}
\ee
where the minus signs have been included in order to match with our conventions on the Lie-algebra relations $d\sigma_1 = - \sigma_2 \wedge \sigma_3$, etcetera.
Notice that 
\bea
\gamma_i & =    i \mathrm{Tr}[\tau_i a_3^{-1}da_3] = \tilde \sigma_i + M_{ij}\Sigma_j =  M_{ij}(\Sigma_j - \sigma_j)  ~.
\eea
We also define
\be
\begin{aligned}
da_1^2 & =  -2 \sum_i (\mathrm{Tr}[\tau_i a_1^{-1}da_1])^2 \\ 
da_2^2 & =  -2 \sum_i( \mathrm{Tr}[\tau_i a_2 da_2^{-1}])^2 \\
da_3^2 & =  -2 \sum_i( \mathrm{Tr}[\tau_i a_3^{-1} da_3])^2 
\end{aligned}
\ee

\section{Derivation of the BPS equations}
\label{tedious}

In the following we derive the BPS system \eqref{eq:BPSwithf} from the $G_2$ structure equations \eqref{torsional}. 
Recall  the metric ansatz is
\be
	ds^2_7  = M \big[ dt^2 + a^2\sum_{i=1}^3 \sigma_i^2  + b^2\sum_{i=1}^3 (\Sigma_i -\tfrac{1}{2}(1+\omega) \sigma_i)^2 \big]~.
\ee
In this section we define the orthonormal frame with an extra factor of $\sqrt{M}$ with respect to the definition \eqref{sframe}, namely here
\bea
e^t  =  \sqrt{M} dt  \qquad~~~ \tilde e^a  = \sqrt{M} a\,  \sigma_a \qquad~~~  e^a =   \sqrt{M} b (\Sigma_a  -\tfrac{1}{2}(1+\omega) \sigma_a)
\eea
so that the associative three form in the following is defined as
\be
\phi    =   e^t \wedge J + \mathrm{Re} [e^{i\theta} \Omega]
\ee
and we use the definitions in \eqref{eq:SU3forms}. Let us look at the first equation in \eqref{torsional}.
 We first compute some useful intermediate results:
\bea
\sqrt{M} d J & = & \frac{d}{d t} \log (ab) e^t \wedge J + 3\frac{b}{4a^2}(1-\omega^2)\frac{1}{3!} \epsilon_{abc} \tilde e^a \tilde e^b \tilde e^c \nn\\
& - & \frac{\omega}{a}\frac{1}{2!} \epsilon_{abc} \tilde e^a \tilde e^b e^c - \frac{1}{b} \frac{1}{2!} \epsilon_{abc} e^a e^b \tilde e^c\\[2mm]
\sqrt{M} \frac{1}{2}d( J\wedge J ) & = & \frac{d}{dt} \log (ab) e^t \wedge J \wedge J\\[2mm]
\sqrt{M} d \mathrm{Re} [\Omega] & = & \frac{d}{dt}\log (b^3)e^t \frac{1}{3!} \epsilon_{abc}  e^a  e^b  e^c + 3\frac{b \omega'}{2a} e^t \frac{1}{3!} \epsilon_{abc} \tilde e^a \tilde e^b \tilde e^c\nn \\
&& -\frac{d}{dt}\log (a^2b)e^t \frac{1}{2!} \epsilon_{abc} \tilde e^a \tilde e^b e^c - \frac{b \omega'}{2a} e^t \frac{1}{2!} \epsilon_{abc} e^a e^b \tilde e^c\nn\\
&& - \frac{1}{2}\left( \frac{1}{b} + \frac{b}{4a^2}(1-\omega^2) \right) J \wedge J\\[2mm]
\sqrt{M} d \mathrm{Im} [\Omega] & = & - \frac{d}{dt} \log a^3 e^t \frac{1}{3!} \epsilon_{abc} \tilde e^a \tilde e^b \tilde e^c + \frac{d}{dt} \log (ab^2) e^t\frac{1}{2!} \epsilon_{abc} e^a e^b \tilde e^c\nn\\
&&  - 2 \frac{b \omega'}{2a} e^t \frac{1}{2!} \epsilon_{abc} e^a \tilde e^b \tilde e^c- \frac{\omega}{2a} J \wedge J
\eea
where we used the identity
\bea
e^1 e^2 \tilde e^1 \tilde e^2 + \mathrm{cyclic} & = & -\frac{1}{2} J \wedge J~.
\eea
After some more algebra we  find
\bea
\sqrt{M} \phi \wedge d \phi & = & e^t \wedge J \wedge J \wedge J \bigg[ \frac{b}{2a} \omega'  + \frac{2}{3} \theta' + \sin\theta \frac{\omega}{a} \nn\\[2mm]
 && ~~~~~~~~~~~~~~~~~ - \cos \theta   \left( \frac{1}{b} + \frac{b}{4a^2} (1-\omega^2) \right) \bigg]~.
\label{pdpeq}
\eea
Thus the first equation  in \eqref{torsional} implies
\be
	\frac{b}{2a} \omega'  + \frac{2}{3} \theta' + \sin\theta \frac{\omega}{a}- \cos \theta   \left( \frac{1}{b} + \frac{b}{4a^2} (1-\omega^2) \right) =0
\ee
Let us now look at the second equation in \eqref{torsional}. We first calculate
\bea
\sqrt{M} d *_7  \phi & = & e^t \wedge \frac{1}{2} J \wedge J \bigg[ \frac{d}{dt}\log (ab)^2  
-\cos \theta \frac{\omega}{a}  \nn\\[2mm]
&& ~~~~~~~~~~~~~~~~~~~ - \sin\theta   \left( \frac{1}{b} + \frac{b}{4a^2} (1-\omega^2) \right)  \bigg]~.
\label{dstarp}
\eea
The equation $d (e^{-2\Phi} *_7 \phi)  =  0$ may be written as
\bea
d*_7 \phi  =  2 d\Phi \wedge *_7 \phi
\eea
and after writing $d\Phi = M^{-1/2} \Phi' e^t $, it may be
regarded as giving the derivative of the dilaton in terms of the remaining functions. In particular
\bea
2 \Phi'  =  \frac{d}{dt}\log a^2b^2   -\cos \theta \frac{\omega}{a} - \sin\theta   \left( \frac{1}{b} + \frac{b}{4a^2} (1-\omega^2) \right)~.
\eea
Finally we analyse the last equation in \eqref{torsional}. This can be rewritten as
\bea
*_7 H_3 = 2 d \Phi \wedge \phi - d\phi = 2 M^{-1/2} \Phi' e^t \wedge \left(\cos\theta  \mathrm{Re} [\Omega] - \sin\theta \mathrm{Im} [\Omega]\right)  - d \phi  
\eea
Then we compute $d\phi$:
\bea
\sqrt{M} d\phi & = & e^t \frac{1}{3!} \epsilon_{abc} e^a e^b e^c \bigg[  \frac{d}{dt} (\cos\theta ) + \cos \theta  \frac{d}{dt} \log b^3  \bigg]+\nn\\[2mm]
&& + e^t \frac{1}{3!} \epsilon_{abc} \tilde e^a \tilde e^b \tilde e^c 
\bigg[ \frac{d}{dt} (\sin\theta ) - 3\frac{b}{4a^2}(1-\omega^2) + 3 \cos \theta \frac{b}{2a}\omega' + \sin\theta \frac{d}{dt} \log a^3 \bigg]\nn\\[2mm]
&& + e^t \frac{1}{2!} \epsilon_{abc} \tilde e^a \tilde e^b e^c \bigg[ -\frac{d}{dt} (\cos\theta ) +\frac{\omega}{a}+ 2\sin\theta \frac{b}{2a} \omega' - \cos \theta \frac{d}{dt} \log (a^2b) \bigg]\nn\\[2mm]
&& + e^t \frac{1}{2!} \epsilon_{abc}  e^a  e^b \tilde e^c \bigg[ -\frac{d}{dt} (\sin\theta ) +\frac{1}{b}
- \cos\theta \frac{b}{2a} \omega' - \sin \theta \frac{d}{dt} \log (ab^2) \bigg]\nn\\[2mm]
& & + \frac{1}{2}J \wedge J \bigg[  \sin\theta \frac{\omega}{a} -\cos\theta  \Big( \frac{1}{b} + \frac{b}{4a^2}(1-\omega^2) \Big)\bigg]~.
\eea
We now need to compute  $*_7 H_3$. First, starting from the ansatz \eqref{fluxansatz} and using 
the relations \eqref{defalfa} we obtain:
\bea
4\sqrt{M} H_3 & = &  - \frac{1}{3!} \epsilon_{abc} e^a e^b e^c \frac{k_2}{b^3}\nn\\[2mm]
&& +  \frac{1}{3!} \epsilon_{abc} \tilde e^a \tilde e^b \tilde e^c \frac{1}{8a^3} \bigg[ 4k_1-k_2 \omega (3+\omega^2) - 3\gamma(1-\omega^2) \bigg]\nn\\[2mm]
&& +  \frac{1}{2!} \epsilon_{abc} \tilde e^a \tilde e^b e^c \frac{1}{4a^2b}\bigg[ - k_2 (1+\omega^2) + 2 \omega \gamma \bigg]\nn\\[2mm]
&& -  \frac{1}{2!} \epsilon_{abc}  e^a  e^b \tilde e^c \frac{1}{2ab^2} (k_2 \omega - \gamma) \nn\\[2mm]
& & - e^t \wedge J \frac{ \gamma' }{2ab} 
\eea
Then the Hodge dual is:
\bea
4\sqrt{M} *_7 H_3 & = & -  e^t \frac{1}{3!} \epsilon_{abc} e^a  e^b  e^c \frac{1}{8a^3}
\Big[  4k_1-k_2 \omega (3+\omega^2) - 3\gamma(1-\omega^2) \Big]\nn\\[2mm]
&& - e^t \frac{1}{3!} \epsilon_{abc} \tilde e^a \tilde e^b \tilde e^c \frac{k_2}{b^3}\nn\\[2mm]
&& +  e^t \frac{1}{2!} \epsilon_{abc} \tilde  e^a \tilde  e^b  e^c \frac{1}{2ab^2} (k_2 \omega - \gamma) \nn\\[2mm]
&& +  e^t \frac{1}{2!} \epsilon_{abc}  e^a e^b \tilde e^c \frac{1}{4a^2b}\Big[ - k_2 (1+\omega^2) + 2 \omega \gamma \Big]\nn\\[2mm]
&& - \frac{1}{2} J \wedge J  \frac{ \gamma' }{2ab}~.
\eea
Putting everything together we find:
\be
	\begin{aligned}
		a' &= \frac{16 a b^3 \sin \theta \big(1 - \omega^2 \big) + 4 a^2 \cos \theta \big( \gamma - k_2 \omega \big) + 4 a b \sin \theta \big(k_2 + k_2 \omega^2 -2 \omega \gamma \big)}{64 a^2 b^2} \\
		&\quad -\frac{b^2 \cos \theta \big( 4 k_1 - 3 k_2 \omega - 32 a^2 \omega - k_2 \omega^3 - 3 \gamma + 3 \omega^2 \gamma \big)}{64 a^2 b^2} \\[2mm]
		b' &= \frac{32 a b^3 \sin \theta \big( \omega^2 - 1 \big) + 4 a^2 \big(\cos \theta - 3 \sec \theta \big) \big(k_2 \omega - \gamma \big)}{64 a^3 b} \\
		&\quad - \frac{4 a b \sin \theta \big(16 a^2 +  k_2 + k_2 \omega^2 - 2 \omega \gamma \big)}{64 a^3 b} \\
		&\quad +\frac{b^2 \sin \theta \tan \theta \big(-4 k_1 + 3 k_2 \omega + 96 a^2 \omega + k_2 \omega^3 + 3 \gamma - 3 \omega^2 \gamma \big)}{64 a^3 b}\\[2mm]
\omega' &= \frac{32 a^3 b \cos \theta + 4 a^2 \sin \theta \big(k_2 \omega - \gamma -16 b^2 \omega \big)}{16 a^2 b^3}  \\
		&\quad + \frac{b^2 \sin \theta \big(4 k_1 - 3 k_2 \omega - k_2 \omega^3 - 3 \gamma + 3 \omega^2 \gamma \big)}{16 a^2 b^3} \\
		&\quad + \frac{4 a b \cos \theta \big( -6 b^2 (\omega^2 - 1) + k_2+ k_2 \omega^2 - 2 \omega \gamma \big)}{16 a^2 b^3} \\[2mm]
    \gamma' &= \frac{- 8 a ^2 \cos \theta + 8 a b \sin \theta \omega + 2 b^2 \cos \theta \big(\omega^2 -1 \big)}{a} \\
	\end{aligned}
\ee
where the angle $\theta$ is fixed in terms of the other functions and reads:
\be
	\cot \theta = \frac{b \Big(12 a^2 \big( \gamma - k_2 \omega \big) + b^2 \big(-4 k_1 + 3 k_2 \omega + 96 a^2 \omega + k_2 \omega^3 + 3 \gamma - 3 \omega^2 \gamma \big) \Big)}{2 a \Big(-4 a^2 \big(k_2 -12 b^2 \big) + 3 b^2 \big(4 b^2 (1 - \omega^2) + k_2 + k_2 \omega^2 - 2 \omega \gamma \big) \Big)}
\ee
Finally,  substituting the functions $a$, $b$ and $\omega$ with the functions $f_i$, and using a computer program to simplify the expressions, 
we find the BPS system \eqref{eq:BPSwithf}.

\section{Supersymmetry conditions in Type IIA}
\label{paperzeroeq}

\subsection{Reduction  from $d=11$}

General conditions characterising ${\cal N}=1$ solutions of eleven-dimensional supergravity of the warped product type $X_{1+2}\times_w M_8$ where $X_{1+2}$ is either $\R^{1,2}$ or AdS$_3$,
were presented in \cite{Martelli:2003ki}.  Here we are interested in the case that $X_{1+2}=\R^{1,2}$.
The eleven-dimensional  metric is written as
\be
	d\hat{s}_{11}^2  =  e^{2\Delta} (d x^2_{1+2}+ds^2_8 )
\ee
and the four-form flux reads
\be
	G  =  e^{3\Delta} (F+ \vol_3 \wedge f)~.
\ee
Thus $F$ is a four-form and $f$ is a one-form. Upon setting $m=0$, the
equations (3.11) - (3.16) of \cite{Martelli:2003ki} become
\be \label{dphieq}
	\begin{aligned}
		d (e^{3\Delta} K \cos \zeta) &= 0 \\
		K \wedge d (e^{6\Delta} *_7 \phi) &= 0 \\
		d (e^{12\Delta} \vol_7 \cos\zeta) &= 0\\
		d\phi \wedge \phi \cos\zeta &= 2 * \left(\cos \zeta f - 2 d\zeta \right) 
	\end{aligned}
\ee
Here  $\phi$ is a three-form, $K$ is a one-form and $\zeta$ is a function, defined as spinor bilinears, 
 that   characterise the $G_2$ structure in eight dimensions. The seven-dimensional Hodge star operator is defined as $*_7 = i_K * $ and $\vol_7 = \tfrac{1}{7} \phi \wedge *_7 \phi$. The electric and magnetic fluxes are the determined in terms of the $G_2$ structure as
\be
	\begin{aligned}
 e^{ -3\Delta} d(e^{3\Delta} \sin\zeta ) & =  f \\
 e^{ -6\Delta} d(e^{6\Delta} \cos\zeta \phi) & =  - * F + \sin\zeta F  
 	\end{aligned}
\ee
The latter equation obeyed by the magnetic flux $F$ may be inverted giving
\be
 \cos^2  \zeta F  =  - e^{ -6\Delta} \left[  \sin \zeta d(e^{6\Delta} \cos\zeta \phi) + * d(e^{6\Delta} \cos\zeta \phi)\right] ~.
\ee
The one-form $K$ in general does not correspond to a Killing vector.  
However, in order to reduce to Type IIA, we will \emph{assume} that the dual vector $K^\#$ is Killing.
In particular,  writing $K = e^{2\Phi/3-\Delta} dy$,  the eleven-dimensional metric takes the form
\be
	d\hat{s}_{11}^2 = e^{2\Delta} (d x^2_{1+2}+ ds^2_7+ e^{4\Phi/3 - 2\Delta} dy^2  )
\ee
and its reduction to ten dimensions then can be simply read off:
\be
	ds_{str}^2 = e^{2\Delta+2\Phi/3} (d x^2_{1+2}+ ds^2_7)~.
\ee
Then we write
\bea
d ~= ~d_7 + dy \de_y~, ~~~~~~~~ f ~= ~ f_7 + dy f_y~.
\eea
Looking first at the fluxes we find
\be
	\begin{aligned}
		f_y &= 0 ~,\\
		f_7 &= e^{ -3\Delta} d_7 (e^{3\Delta} \sin\zeta)~.
	\end{aligned}
\ee
Using these  equations, from \eqref{dphieq} we obtain
\be 
	\begin{aligned}
			d_7 \big( e^{6\Delta} *_7 \phi \big) &= 0 \\
		\phi \wedge d_7 \phi &= 0 \\
		2 d_7 \zeta - e^{-3\Delta} \cos \zeta d_7 \big( e^{3\Delta} \sin \zeta \big) &= 0\\
		d_7 \big( e^{2\Delta+ 2\Phi/3} \cos \zeta \big) &= 0 
	\end{aligned}
\ee
Finally, the reduction of the four-form $G$ gives the NS three-form $H_3$ and the RR four-form $F_{4}$:
\be
	\begin{aligned}
		H_3 &= \frac{1}{\cos^2 \zeta} e^{-4\Delta + 2\Phi/3} *_7 d_7 \big( e^{6\Delta} \cos \zeta \phi \big)~, \\
		F_{4} &= \vol_3 \wedge d_7 \big( e^{3\Delta} \sin \zeta \big) - \frac{\sin \zeta}{\cos^2 \zeta} e^{-3\Delta} d_7 \big(e^{6\Delta} \cos \zeta \phi \big)~.
	\end{aligned}
\ee

\subsection{Killing spinor ansatze}

In this Appendix we discuss ansatze for the Killing spinors in eleven and ten dimensions. 
Although in the main text we have not derived the equations from the Killing spinors directly, it may be
useful to spell out some details about spinors and representations of gamma matrices. 
The general spinor ansatz in eleven dimensions reads 
 \cite{Martelli:2003ki}
\be
\eta = e^{\Delta/2} \psi \otimes \xi  = e^{\Delta/2} \psi \otimes (\xi_+  + \xi_-)~.
\label{elevendspinor}
\ee
We use the following explicit  representation of gamma matrices 
\be
\begin{aligned}
	\hat{\Gamma}^\mu & =  -e^{-\Delta}  \rho^\mu \otimes  \hat \gamma^\chi \qquad~~ \mu =0,\dots,2\\
	\hat{\Gamma}^m & =  e^{-\Delta}  \mathbf{1} \otimes \hat \gamma^m \qquad~~ ~~m =3,\dots ,10
\end{aligned}
\ee
The $\hat \gamma^m$ are $16\times 16$  gamma-matrices  and $\hat \gamma^\chi $ is the chirality matrix in $d=8$,  with $(\hat \gamma^\chi)^2= \mathbf{1}$. 
The  $\rho^\mu$ denote $2\times 2$ gamma-matrices in $d=1+2$. In an explicit representation these
may be taken  \cite{Martelli:2003ki} as follows
\bea
	&& \rho^0 = i \sigma^1~, \quad \rho^1 = \sigma^2~, \quad \rho^2 = \sigma^3 ~,
\eea
where $\sigma^i$ are Pauli matrices. 
The Majorana condition in eleven dimensions  $\eta= \eta^c = D_{11}\eta^*$, with $D_{11}=\sigma^3 \otimes \mathbf{1}$ implies that $\psi^* = \sigma^3 \psi$ and   $\xi_\pm =\xi^*_\pm$. 
Hence $\psi $ is a Majorana spinor in $d=1+2$ and $\xi_\pm$ are 
Majorana-Weyl spinors in $d=8$.

To make the reduction to ten dimensions it is convenient to  chose the following 
representation of $d=8$ gamma-matrices:
\be\begin{aligned}
	\hat \gamma^m = \left\{ \begin{array}{rcl}
	\hat \gamma^i & = & \sigma^2 \otimes \gamma^i \qquad i =3,\dots ,9 \\
	\hat \gamma^{10} & = & \sigma^3 \otimes \mathbf{1}
\end{array}\right.
\end{aligned}
\ee
These are real and symmetric, taking $\gamma^i$ to be purely imaginary and anti-symmetric. 
Then the $d=8$ chirality matrix is 
\be
\hat \gamma^\chi = \hat \gamma^3 \cdots \hat \gamma^{10} = - \sigma^1 \otimes  \mathbf{1}
\ee
and is again real and symmetric. We then write the $d=8$ Majorana-Weyl spinors as
\be
\xi_+ = \alpha_+ \otimes \beta_+~, \qquad ~~~~~ \xi_- = \alpha_- \otimes \beta_-~,
\ee
where $\alpha_\pm$  are two-component spinors and $\beta_\pm$ are eight-component spinors, which can all be taken to be real. 
Imposing the  $d=8$ chirality conditions $\hat \gamma^\chi \xi_\pm   =  \pm \xi_\pm $
then implies 
\bea
\sigma^1 \alpha_\pm = \mp \alpha_\pm ~.
\eea
Thus, up to an overall real function, we can take 
\bea
\alpha_+ = \binom{1}{-1}~, ~~~~~~~~~ \alpha_- = \binom{1}{1}~.
\eea
Upon reducing to Type IIA, the complexified Killing spinor is simply related to the Killing spinor in eleven dimension as 
$\eta = e^{-\Phi/6} \epsilon$, hence the ansatz may be written as  
\bea
\epsilon = \epsilon_1 + \epsilon_2 = e^{\Phi/6+\Delta/2} \psi \otimes \left( \theta_1 \otimes \chi_1 + \theta_2 \otimes \chi_2\right) ~.
\label{tendspinor}
\eea
The  ten-dimensional gamma matrices are the same as the eleven-dimensional ones up to a warp factor, where the eleventh one becomes the ten-dimensional chirality matrix.  Namely
\be
\begin{aligned}
\Gamma^\mu & =  e^{-\Delta-\Phi/3} \rho^\mu \otimes  \sigma^1 \otimes  \mathbf{1} \qquad ~~\mu  = 0,\dots ,2\\
\Gamma^i & =  e^{-\Delta-\Phi/3}  \mathbf{1} \otimes \sigma^2 \otimes \gamma^i \qquad ~~i =3,\dots,9 \label{newtenrep}
\end{aligned}
\ee
and the ten-dimensional chirality matrix is 
\bea
\Gamma^\chi & = & \mathbf{1} \otimes \sigma^3 \otimes \mathbf{1} 
\eea
where we have used the convention that\footnote{Taking $\gamma^3 \gamma^4 \cdots \gamma^{9} = - i$ gives 
$\Gamma^\chi  = - \mathbf{1} \otimes \sigma^3 \otimes \mathbf{1}$. } 
\bea
\gamma^3 \gamma^4 \cdots \gamma^{9} = + i ~.
\eea
The Majorana condition on the complexified ten-dimensional spinor 
is $\epsilon = \tilde D_{10}\epsilon^*$, where we can take $\tilde D_{10} = \sigma^3 \otimes \mathbf{1} \otimes \mathbf{1}$. Therefore  we have 
\be
\begin{aligned}
\psi^* \otimes \theta_1^* \otimes \chi_1^* & = \sigma^3\psi \otimes  \theta_1 \otimes \chi_1 \\
\psi^* \otimes \theta_2^* \otimes \chi_2^* & =  \sigma^3\psi \otimes \theta_2 \otimes \chi_2 
\end{aligned}
\ee
which can be solved for example taking
\bea
\sigma^3\psi = \psi^* ~,\qquad ~~~~\theta_i =\theta_i^*~, \qquad~~~~ \chi_i = \chi_i^*~.
\eea
The ten-dimensional chirality conditions give
\be
\begin{aligned}
\Gamma^\chi \epsilon_1 = \epsilon_1 & \Rightarrow   \sigma^3 \theta_1 = \theta_1  \\
\Gamma^\chi \epsilon_2 = -\epsilon_2 & \Rightarrow   \sigma^3 \theta_2 = -\theta_2
\end{aligned}
\ee
Therefore, up to overall factors (not necessarily constant) that we can reabsorb in $\chi_i$, we can 
take 
\bea
\theta_1 = \binom{1}{0}~, ~~~~~~~~~~ \theta_2 = \binom{0}{1}~.
\eea
Comparing (\ref{elevendspinor}) with (\ref{tendspinor}) we find
\be
\begin{aligned}
\xi_+ & = \frac{1}{2} \binom{1}{-1} \otimes (\chi_1 - \chi_2) \\
\xi_- & = \frac{1}{2} \binom{1}{1} \otimes (\chi_1 + \chi_2) 
\end{aligned}
\ee
Now using 
\be
\begin{aligned}
\xi_+^T \xi_+ + \xi_-^T \xi_- & =  2 \\
\xi_+^T \xi_+ -  \xi_-^T \xi_- & =  2\sin \zeta
\end{aligned}
\ee
from  \cite{Martelli:2003ki}, we compute
\be
\begin{aligned}
\chi_1^T \chi_1 + \chi_2^T \chi_2 & =  2\\
\chi_1^T \chi_2 & =  - \sin\zeta
\end{aligned}
\ee
Since we are interested in a strict $G_2$ structure, we must have $\chi_1 \propto \chi_2$. In fact, if $\chi_1$ and $\chi_2$ were not parallel, we could construct a vector $\chi_1^T\gamma_i \chi_2$, reducing the structure to $SU(3)$. Then we have that
\be
\begin{aligned}
\chi_1 & =  \sqrt{2}\sin\alpha \, \chi \nonumber\\
\chi_2 & =  \sqrt{2}\cos\alpha \, \chi
\end{aligned}
\ee
where we normalised the spinor $\chi$ as $\chi^T \chi =1$ and 
$\sin 2\alpha  =  - \sin\zeta$.
In conclusion, we have shown that the ten-dimensional Killing spinor takes the form 
\bea
\epsilon & = & e^{\Phi/6+\Delta/2} \psi \otimes \sqrt{2} \binom{\sin \alpha}{\cos \alpha} \otimes \chi ~. 
\eea
Note that this possibility is excluded in the spinor ansatz in \cite{Haack:2009jg}. 
See \emph{e.g.} equation (A.3) of this reference.

\end{document}